\documentclass[%
 aip,
 amsmath,amssymb,
 reprint,%
]{revtex4-1}

\usepackage{graphicx}
\usepackage{dcolumn}
\usepackage{bm}
\usepackage[normalem]{ulem}
\usepackage[utf8]{inputenc}
\usepackage[T1]{fontenc}
\usepackage{mathptmx}
\usepackage{color}

\newcommand{\ii}{\rm i}

\begin{document}

\preprint{AIP/123-QED}

\title[From single-particle-like to interaction-mediated plasmonic resonances in graphene nanoantennas]{From single-particle-like to interaction-mediated plasmonic resonances in graphene nanoantennas}


\author{Marvin M. M\"uller}
\email{marvin.mueller@kit.edu}
\affiliation{Institute of Theoretical Solid State Physics, Karlsruhe Institute of Technology (KIT), 76131 Karlsruhe, Germany}
\author{Miriam Kosik}
\email{mkosik@doktorant.umk.pl}
\affiliation{Institute of Physics, Nicolaus Copernicus University in Toru\'n, Grudziadzka 5, 87-100 Toru\'n, Poland}
\author{Marta Pelc}
\affiliation{Institute of Physics, Nicolaus Copernicus University in Toru\'n, Grudziadzka 5, 87-100 Toru\'n, Poland}
\affiliation{Donostia International Physics Center (DIPC), Paseo Manuel Lardizabal 4, 20018 Donostia-San Sebasti\'an, Spain}
\affiliation{Centro de F\'isica de Materiales, CFM-MPC CSIC-UPV/EHU, Paseo Manuel Lardizabal 5, 20018 Donostia-San Sebasti\'an, Spain}
\author{Garnett W. Bryant}
\affiliation{Joint Quantum Institute, University of Maryland and National Institute of Standards and Technology, College Park, Maryland 20742, USA}
\affiliation{Nanoscale Device Characterization Division, National Institute of Standards and Technology, Gaithersburg, Maryland 20899, USA}
\author{Andr\'es Ayuela}
\affiliation{Donostia International Physics Center (DIPC), Paseo Manuel Lardizabal 4, 20018 Donostia-San Sebasti\'an, Spain}
\affiliation{Centro de F\'isica de Materiales, CFM-MPC CSIC-UPV/EHU, Paseo Manuel Lardizabal 5, 20018 Donostia-San Sebasti\'an, Spain}
\author{Carsten Rockstuhl}
\affiliation{Institute of Theoretical Solid State Physics, Karlsruhe Institute of Technology (KIT), 76131 Karlsruhe, Germany}
\affiliation{Institute of Nanotechnology, Karlsruhe Institute of Technology (KIT), 76021 Karlsruhe, Germany}
\author{Karolina S\l{}owik}
\affiliation{Institute of Physics, Nicolaus Copernicus University in Toru\'n, Grudziadzka 5, 87-100 Toru\'n, Poland}



\date{\today}

\begin{abstract}
Plasmonic nanostructures attract tremendous attention as they confine electromagnetic fields well below the diffraction limit while simultaneously sustaining extreme local field enhancements. To fully exploit these properties, the identification and classification of resonances in such nanostructures is crucial. Recently, a novel figure of merit for resonance classification has been proposed~\cite{Mueller2020} and its applicability was demonstrated mostly to toy model systems. This novel measure, the energy-based plasmonicity index (EPI), characterizes the nature of resonances in molecular nanostructures. The EPI distinguishes between either a single-particle-like or a plasmonic nature of resonances based on the energy space coherence dynamics of the excitation. To advance the further development of this newly established measure, we present here its exemplary application to characterize the resonances of graphene nanoantennas. In particular, we focus on resonances in a doped nanoantenna. The structure is of interest, as a consideration of the electron dynamics in real space might suggest a plasmonic nature of selected resonances in the low doping limit but our analysis reveals the opposite. We find that in the undoped and moderately doped nanoantenna, the EPI classifies all emerging resonances as predominantly single-particle-like and only after doping the structure heavily, the EPI observes plasmonic response.

\end{abstract}

\maketitle

\section{\label{sec:introduction}Introduction}

The field of plasmonics experienced huge interest in the last two decades ~\cite{Cazalilla2000,Maier2007,Pelton2008,Giannini2011,Koppens2011} with a recent shift of focus towards related quantum processes and applications at the nanoscale~\cite{Tame2013, Bryant2014, Bolzhevolnyi2017}. The possibility of confining and enhancing electromagnetic fields \cite{Takahara1997, Gramotnev2010,Novotny2012, Giannini2011} attracts attention not only for fundamental research reasons but also due to many potential applications in plasmonic sensing~\cite{Melendez1996, Awazu2007, Lee2016}, photodetection\cite{Kim2014,Brongersma2015,Zhang2015,Yu2016}, medicine~\cite{Hirsch2003,Qian2008}, optical metamaterials\cite{Ju2011,Hess2012,Muhlig2013} and single photon sources\cite{Koenderink2009,Chen2010}. \par 
Graphene supports intrinsic tunable plasmons and, therefore, is a well-suited platform for exploring and exploiting plasmonic phenomena~\cite{Ju2011,deAbajo2014,Goncalves2016}. Recent progress in nanostructure fabrication allows to produce graphene flakes consisting of only a few hundred atoms~\cite{Yamada2013,Jabari2019} that support a plasmonic response at near infrared frequencies. Since a classical description based on the Drude model fails to properly predict the properties of graphene nanoantennas with a size below $10\,\rm nm$~\cite{Thongrattanasiri2012,Manjavacas2013}, more accurate attempts to model such systems should account for fine details at atomistic scale and are mostly based on quantum mechanical methods. Among these one can find density functional theory (DFT)~\cite{Piccini2013,Bursi2016,Noguchi2005,Manjavacas2013TDDFT}, the tight binding (TB) model~\cite{Ezawa2007,Gucclu2010,Jaskolski2011,Cox2014}, or quantum fluid dynamics~\cite{Reinhard1990,Raitza2012}.  \par 

Graphene nanoantennas support both single-particle-like resonances as well as plasmonic ones, which leads to the important question how to identify the nature of resonances in such nanostructures~\cite{Townsend2014,Zhang2017}. This complex issue has received substantial interest in the last years and several works have been devoted to address it. The collective charge density oscillation in real space is typically considered as the smoking gun to classify a specific resonance as plasmonic in nature. However, as will be shown below, the real space analysis cannot be the sole basis for decisions on the nature of the resonance. Therefore, it continues to be a major challenge in the field of plasmonics to decide whether a specific excitation is plasmonic or not. Associated to that charaterization is, of course, the question how to actually define a plasmonic excitation in nanoscaled systems. This contribution addresses these questions and aims to provide an answer. \par
Early studies found that in graphene nanoantennas a single extra electron from doping can switch on infrared plasmons that were absent from the structure before doping and that adding further electrons causes a significant shift in plasmon frequency~\cite{Manjavacas2013}. Studying how the spectral position of a resonance changes with addition of doping electrons is one of the clues that can serve as a guide to determine its character. 

\begin{figure*}
\includegraphics[width=0.75\linewidth]{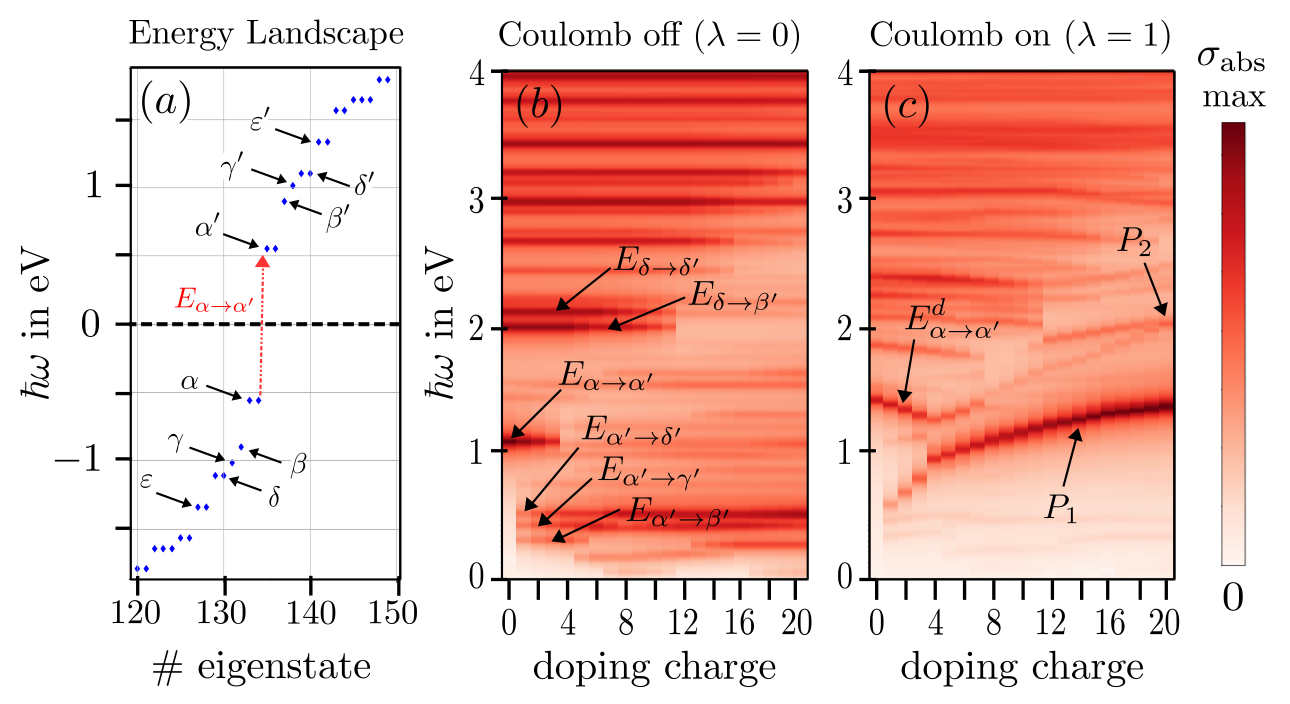}
\caption{\label{fig:coulomb_on_off} a) Jab\l{}onski diagram of the armchair triangle consisting of 270 atoms. Only the neighborhood of the HOMO-LUMO gap is shown, which determines the properties of resonances discussed in the main text. b) Absorption cross section of the graphene armchair triangle consisting of 270 atoms as a function of doping charge. Coulomb interaction is turned off ($\lambda$ = 0). c) Absorption cross section of the graphene armchair triangle consisting of 270 atoms as a function of doping charge. The Coulomb interaction is fully taken into account ($\lambda = 1$).}
\end{figure*}

Another approach to this problem relies on scaling of the electron-electron interaction strength in simulations to find how the frequency of a given resonance depends on this variation~\cite{Bernadotte2013,Krauter2015,Piccini2013}. In this approach, the excitations that show little dependence on the scaling parameter are classified as single-particle like, while those which blue shift considerably with the increase of the Coulomb interaction strength are deemed plasmonic. \par
Bryant and Townsend have used real space and real-time time-dependent DFT to examine jellium spheres. They were able to distinguish two different types of behavior that the occupations of electronic states in energy basis of a given resonance can show. The first one they called "sloshing". It is a pattern of oscillatory movement between shells above and below the Fermi energy. The second they called "inversion", which is associated with a continuous transition of electrons from occupied to unoccupied states~\cite{Townsend2014, Townsend2015}. The ratio of "sloshing" and "inversion" that is linked to a given resonance can be used as a clue to determine its nature. 
Other attempts to classify resonances were based on a bi-local auto-correlation function~\cite{Raitza2012} or a model in which the electrons were confined to a potential box~\cite{Jain2014}. \par 
Among the numerous studies conducted on this topic, a line of research focused on the construction of a universal figure of merit for resonance classification. In 2016, Bursi \textit{et al.} proposed the “plasmonicity index” (PI) to characterize and quantify plasmonic behavior based on how much the induced potential in a nanostructure deviates from a neutral case. Shortly afterwards, Zhang \textit{et al.} defined a dimensionless, but unnormalized metric called generalized plasmonicity index (GPI) to distinguish plasmons from single-particle-like excitations based on a similar aspect~\cite{Zhang2017}. Both of these measures can be determined using the real space charge distribution of the structure's resonances as the only input. However, as we outline below, there exist resonances of single-particle-like nature that reveal strong dipolar character even in \textit{non-interacting} systems. On the other hand, Pines and Bohm state in their pioneering work in 1952, that charge oscillations may have individual and collective components, where the latter emerge only in systems with long-range electron-electron interaction~\cite{Pines1952}. \par  
As a contribution to resolve this issue, recently, a new figure of merit for resonance classification has been proposed~\cite{Mueller2020}. The energy-based plasmonicity index (EPI) is a normalized and dimensionless measure for characterizing the nature of resonances in nanostructures. It does not rely on charge carrier oscillation patterns on the nanostructure and, hence, cannot be determined by the analysis of atomic site population dynamics. Unlike the PI and the GPI, the EPI probes the manifestation of the resonance directly in energy space. It quantifies if the existence of a given resonance can be explained predominantly by the system's energy landscape or if electronic interaction energy, \textit{e.g.} Coulomb energy, needs to be taken into account to properly determine its spectral position. 
The definition of the EPI is based on the coherences of the system's density operator and the single-particle energies. 
While the EPI has been applied so far mostly to toy model systems, it remains an open question how this measure can be used to explain and to understand resonances sustained in structures of practical relevance. Here, we concentrate on the study of an armchair-edged graphene nanoantenna with triangular shape that is of relevance in the context of nonlinear frequency conversion processes such as higher harmonic generation, for instance~\cite{Cox2014}. \par
The paper is organized as follows. In the Methods section, we shortly introduce the basics of the model and the EPI measure. Further, we present a thorough analysis of a few chosen resonances in a graphene nanoantenna doped with various numbers of electrons. To fully understand the nature of each resonance, we analyze the dependence of the absorption spectra on doping with and without Coulomb interaction taken into account, the dependence of spectral positions and absolute strengths of the nanoantenna's resonances on Coulomb interaction scaling, their real space charge distribution, and energy space fingerprints. Then, we compute and discuss the EPI for the shown resonances. As a conclusion, we find that in the undoped and moderately doped nanoantenna, the EPI classifies all emerging resonances as predominantly single-particle-like. Only in the 20-fold heavily doped nanoantenna, the EPI observes truly plasmonic response. 

\section{\label{sec:method}Method}
Our general modelling framework is based on previous work by Cox and Garcia de Abajo~\cite{Cox2014}. We model the graphene nanostructures relying on the tight binding (TB) approach~\cite{Wallace1947} in the nearest-neighbor approximation. The coupling to an externally applied laser field is considered in the quasistatic limit and dipolar approximation. To describe the Coulomb interaction between electrons at different carbon sites $l$ and $l'$, we employ a Coulomb interaction matrix $v^\lambda_{ll'} = \lambda v_{ll'}$, where $v_{ll'}$ values are based on Ref.~\onlinecite{Potasz2010} for the onsite-, nearest neighbor-, and next-to-nearest neighbors. For the atomic sites that are further away from each other, we use the usual $1/r$ power law. The parameter $\lambda\in[0,1]$ is used to continuously vary the Coulomb interaction from being completely turned off ($\lambda = 0$) to being fully taken into account ($\lambda = 1$). To describe the dynamics of the system, we first construct the ground state density matrix from the TB-Hamiltonian eigenstates of the nanoantenna according to the aufbau principle. Then, we evolve this state with a master equation that contains the influence of an external optical illumination as well as a phenomenological damping term that accounts for the dissipative processes in the nanoantenna.
This method allows to find the optical response of graphene nanoantennas in terms of the time dependent polarization function of the system \cite{Cox2014,Mueller2020} and the absorption spectrum related to its Fourier transform. Different resonances observed in the absorption spectrum can then be related to either single-particle-like or collective effects, and in general both physical mechanisms contribute to a given resonance and determine its spectral position. It is the main challenge to find out which resonance is caused by which effect.  

The figure of merit we will use to quantify the contribution of collective effects to a specific resonance is the energy-based plasmonicity index (EPI). It is based on the stationary density matrix $\rho^\omega$ of a nanoantenna subject to continuous wave (CW) illumination at the resonance frequency $\omega$. To define the EPI, we introduce an auxiliary quantity $\tilde{\rho}^\omega$ related to $\rho^\omega$ as follows:
\begin{align}
\tilde{\rho}_{jj'}^\omega = \frac{|\rho_{jj'}^\omega|}{\left| |E_j-E_{j'}|-\hbar\omega+\ii\epsilon\right|^2},
\label{eq:scaling}
\end{align}
where the elements $\rho^\omega_{jj^\prime}$ of the density matrix are given in the basis of TB-Hamiltonian eigenstates. The parameter $\epsilon=0.05\,\rm eV$ in the denominator prevents a divergence if the incident illumination frequency $\omega$ happens to be perfectly resonant to the transition frequency between the electronic energy states $E_j$ and $E_{j'}$. Since information on electronic transitions that contribute to the density-matrix dynamics is already contained in its off-diagonal elements (coherences), we deplete the diagonal of the density matrix $\rho^\omega_{jj}\rightarrow 0$ before it enters the definition of $\tilde{\rho}$ in Eq.\,\ref{eq:scaling}. Please note again, that in this way, no information about the occupations of energy states enters the definition of the EPI. \par
Having introduced $\tilde{\rho}$, we define the EPI according to
\begin{align}
\label{epi_def}
\rm EPI(\omega) &= 1-\langle\tilde{\rho}^\omega,\rho^\omega\rangle\,\in [0,1],
\end{align}
where we use a scalar product of two matrices $a$ and $b$ which is defined as
\begin{align}
\langle a, b\rangle &:= \frac{\sum_{mn}|a_{mn}b_{mn}|}{(\sum_{mn}|a_{mn}|^2\cdot\sum_{mn}|b_{mn}|^2)^{1/2}}\,\in [0,1].
\end{align}  
For a resonance that comprises the single-particle-like transition from state $|j\rangle$ to $|j'\rangle$, the density operator's coherence element $\rho^\omega_{jj'}$ is non-zero. In the definition of $\tilde{\rho}^\omega_{jj'}$, these elements get enhanced in case the excitation energy $\hbar\omega$ matches the energy difference of the single-particle states, $|E_j-E_{j'}|$. On the other hand, if coherence elements in the density operator $\rho^\omega$ exist, which cannot be related to the excitation energy, they get suppressed in $\tilde{\rho}^\omega$. Consequently, if a resonance is predominantly composed of single-particle-like transitions, we find $\tilde{\rho}^\omega\approx k\cdot\rho^\omega$ with a constant $k$, furthermore $\langle \tilde{\rho}^\omega,\rho^\omega\rangle \approx 1$, and therefore $\text{EPI}(\omega)\approx 0$, which renders the resonance single-particle-like. On the other hand, if a resonance is comprised predominantly by coherences that cannot be associated with the excitation energy (which is the case in plasmonic resonances), we find $\tilde{\rho}^\omega\approx 0$, because most of the coherence elements in $\rho^\omega$ get suppressed, furthermore $\langle \tilde{\rho}^\omega,\rho^\omega\rangle \approx 0$, and therefore $\text{EPI}(\omega)\approx 1$. With that we get an explicit normalized measure of whether a resonance is single-particle-like or plasmonic. 
More details concerning the EPI, especially graphical illustrations of the density operator coherence elements for the single-particle-like and plasmonic resonances, and a thourough analysis of the measure, as well as a complete description of the used methodology, can be found in literature~\cite{Mueller2020}.

\section{\label{sec:analysis}Resonance analysis}
In this section, we examine the absorption spectrum of a triangular armchair-edged graphene nanoantenna consisting of $N = 270$ atoms. The eigenenergies of the non-interacting system that are located near the Fermi energy of an undoped flake $E_F$ = $0$ eV are shown in Fig.~\ref{fig:coulomb_on_off}a. We label the eigenstates corresponding to different energy levels with Greek letters. Note that some of these might be degenerate, e.g. in Fig.~\ref{fig:coulomb_on_off}a the letter $\alpha$ denotes the $133^\mathrm{rd}$ and $134^\mathrm{th}$ eigenstates. \par 

In the absorption spectrum of the considered nanostructure, one can identify a few resonances that correspond to single-particle-like transitions. This can be seen particularly well in Fig.~\ref{fig:coulomb_on_off}b, where the Coulomb interaction and, therefore, collective interaction-mediated processes are turned off. Some prominent resonances are present, e.g. at 1.12 eV (associated with the $\alpha\longrightarrow\alpha^{\prime}$ transition), at 2 eV ($\delta\longrightarrow\beta^{\prime}$ transition) and at 2.2 eV ($\delta\longrightarrow\delta^{\prime}$ transition). Moreover, after doping the system with one electron, which populates the $\alpha^{\prime}$ state and enables transitions from it, three new resonances in the absorption spectrum appear in the energy range between 0.2 eV and 0.5 eV. These clearly correspond to the $\alpha^{\prime}\longrightarrow\beta^{\prime}$,
$\alpha^{\prime}\longrightarrow\gamma^{\prime}$, and $\alpha^{\prime}\longrightarrow\delta^{\prime}$ transitions. Moreover, we observe that the resonance associated with the $\alpha\longrightarrow\alpha^{\prime}$ transition vanishes after doping with four additional electrons, since the $\alpha'$ states are fully occupied and cannot serve as acceptors for any transition any more. \par 
When the Coulomb interaction is taken into account (Fig.~\ref{fig:coulomb_on_off}c), the character of some resonances is modified. 
In particular, some excitations blue shift with increasing number of doping electrons. If they appear only when the interaction between a collection of electrons is allowed, we expect them to originate from electron-electron interaction. We will denote the fundamental mode as $P_1$ and the higher-order mode as $P_2$~(Fig.~\ref{fig:coulomb_on_off}c). Apart from these resonances, we find resonances of a different type. They are mostly of a single-particle character, but are dressed (superscript $d$) by interaction energy, as we argue below. These can still be attributed to a transition between a single pair of eigenstates, e.g. the resonance denoted as $E^d_{\alpha\longrightarrow\alpha^{\prime}}$ in Fig.~\ref{fig:coulomb_on_off}c. While plasmonic resonances are in general known to blue shift with increasing charge carrier density, this specific example of a dressed single-particle-like resonance appears to red shift as additional electrons are introduced into the system. Interestingly, another example of a dressed resonance, the one associated with the $\alpha^{\prime}\longrightarrow\delta^{\prime}$ transition, behaves differently and blue shifts. From this observation and from the similar charge carrier distribution of these two resonances in real space (see Fig.~\ref{fig:real_space_d2} later in the manuscript), we conclude that the shifting behavior of a resonance as a function of charge carrier density cannot be used to unambiguously determine the nature of said resonance. \par
\begin{figure}
\includegraphics[width=\linewidth]{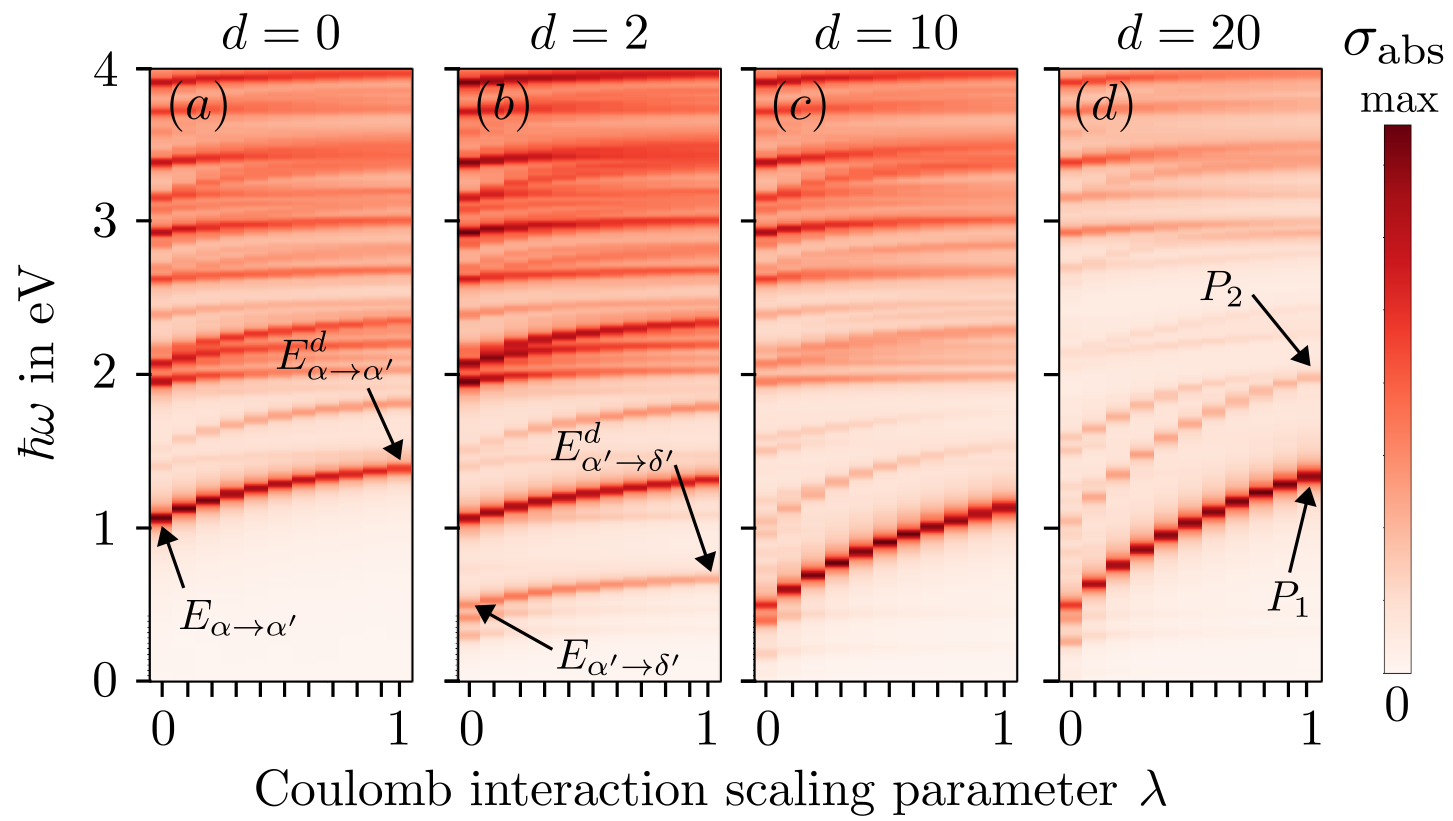}
\caption{\label{fig:absorption_of_lambda} Absorption spectra of the triangular 270-atom graphene nanoantenna as functions of the Coulomb scaling parameter $\lambda$ plotted for four levels of doping ($d$ = 0, 2, 10, 20).}
\end{figure}

\begin{figure}
\includegraphics[width=0.65\linewidth]{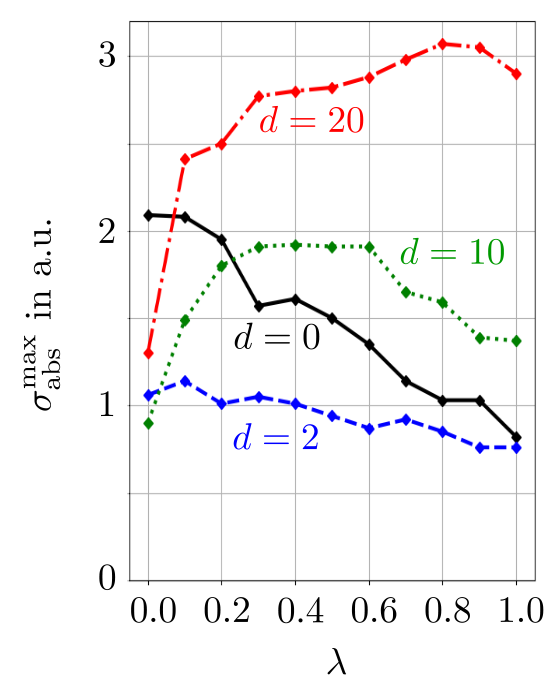}
\caption{\label{fig:max_abs} Strength of the resonances around 1 eV in Fig.~\ref{fig:absorption_of_lambda} as a function of the Coulomb scaling parameter $\lambda$ plotted for four levels of doping ($d$ = 0, 2, 10, 20). $\sigma_{\rm abs}^{\rm max}$ stands for the peak value of the absorption cross section at resonance.}
\end{figure}

 
For a closer inspection, we employ the scaling approach \cite{Bernadotte2013,Krauter2015} and smoothly scale the Coulomb interaction strength by a parameter $\lambda\in\left[0,1\right]$. A look at how the absorption spectrum changes with $\lambda$ shows that in the undoped structure, there is a continuous transition from the $E_{\alpha\longrightarrow\alpha^{\prime}}$ resonance at 1.12 eV for $\lambda = 0$ to the dressed $E^d_{\alpha\longrightarrow\alpha^{\prime}}$ resonance at 1.4 eV for $\lambda = 1$ (Fig.~\ref{fig:absorption_of_lambda}a). Therefore, we can argue that the $E^d_{\alpha\longrightarrow\alpha^{\prime}}$ resonance in Fig.~\ref{fig:coulomb_on_off}c exists also without Coulomb interaction and that it is dominated by the single-particle component, hence it is not predominantly plasmonic. Interestingly, the same resonance in the case of the two-fold doped nanostructure ($d=2$) behaves similarly (Fig.~\ref{fig:absorption_of_lambda}b) which suggests it is also more of a single-particle-like nature. A qualitative difference appears in the scaling of the most prominent resonances in the absorption spectra in the case of 10 or 20 doping electrons (Figs.~\ref{fig:absorption_of_lambda}c and d). A close look at the absolute strength of these resonances highlights the difference between the undoped/two-fold doped and the 10-/20-fold doped cases even more (Fig.~\ref{fig:max_abs}). While the first pair of resonances ($d=0,2$) decreases in strength with increasing value of Coulomb interaction, the latter two behave differently. In the structure with strongest doping ($d = 20$) the relation is exactly opposite and the resonance becomes stronger with growing Coulomb interaction. The resonance in the 10-fold doped structure is an intermediate case that grows with $\lambda$ to a certain point, then decreases. From this, we deduce that there is a qualitative difference between the resonance $P_1$ in the 20-fold doped nanoantenna and the resonance $E^d_{\alpha\longrightarrow\alpha^{\prime}}$ in the two-fold doped structure.

\subsection{Real Space Dynamics}
\begin{figure}[h]
\includegraphics[width=\linewidth]{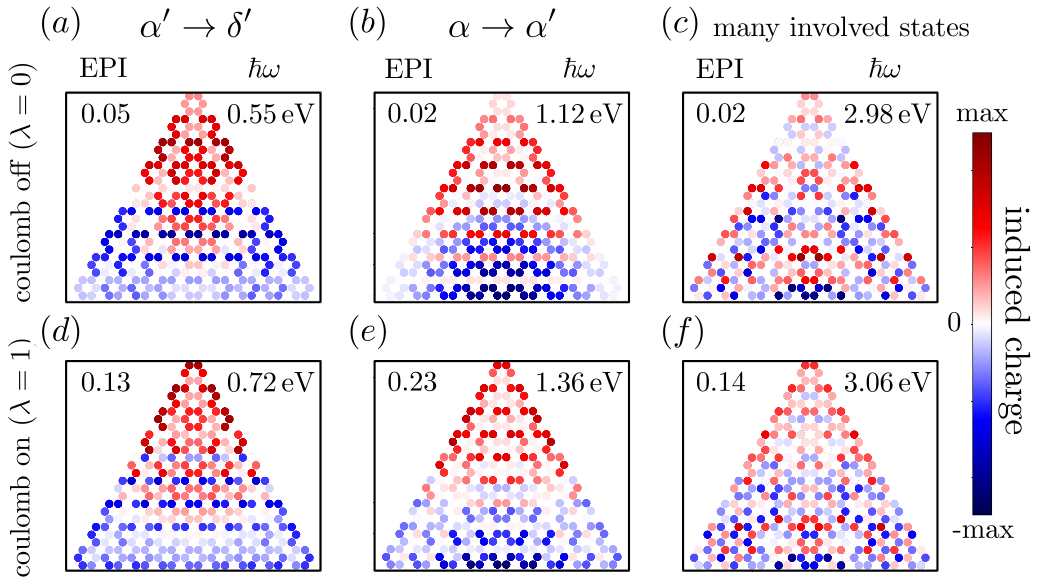}
\caption{\label{fig:real_space_d2} Comparison of the induced charge distribution with Coulomb interaction turned off ($\lambda$ = 0, upper row) and taken into account ($\lambda$ = 1, lower row). Snapshots present the real space induced charge distribution in the triangular 270-atom graphene nanoantenna with two doping electrons under vertically polarized CW illumination. Snapshots were taken at the time of maximum dipole moment. Corresponding values of the EPI are shown in the upper-left corners of the subfigures.}
\end{figure}

Another important feature of a resonance is its real space dynamics. We compare the resonances in a two-fold doped and in a 20-fold doped nanoantenna. In Fig.~\ref{fig:real_space_d2}, the results for a structure with 2 doping electrons are shown. The real space distribution of the induced charge at resonances corresponding to the $\alpha^{\prime}\longrightarrow\delta^{\prime}$ (at 0.55 eV for $\lambda = 0$ and 0.72 eV for $\lambda = 1$) and $\alpha\longrightarrow\alpha^{\prime}$ (at 1.12 eV for $\lambda = 0$ and 1.36 eV for $\lambda = 1$) transitions shows dipolar character (Figs.~\ref{fig:real_space_d2}a, b, d, and e).
This dipolar character can cause some of the real space based plasmonicity metrics to qualify these resonances as plasmonic. The GPI, for instance, exhibits peaks both for the $E^d_{\alpha^\prime\longrightarrow\delta^{\prime}}$ resonance around 0.72 eV and the $E^d_{\alpha\longrightarrow\alpha^{\prime}}$ resonance around 1.4 eV of strengths 3 and 5.5, respectively, for the very same structure in the two-fold doped case~\cite{Zhang2017}. Thus, in the sense of the GPI, they are both classified plasmonic. \par
\begin{figure}[h]
\includegraphics[width=\linewidth]{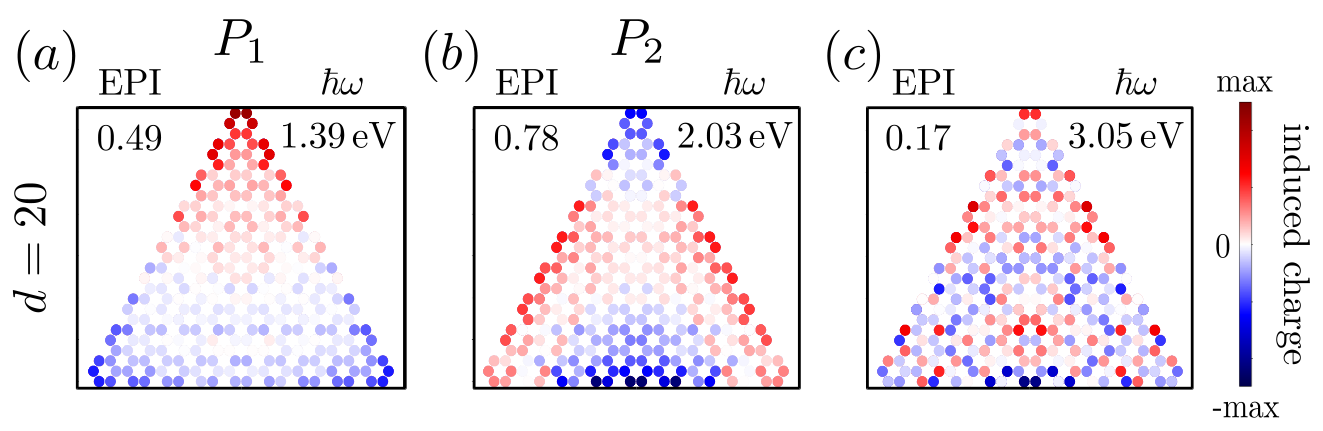}
\caption{\label{fig:real_space_d20} Snapshots present the real space induced charge distribution in the triangular 270-atom graphene nanoantenna with 20 doping electrons under vertically polarized CW illumination. Snapshots were taken at the time of maximum dipole moment. Corresponding values of the EPI are shown in the upper-left corners of the subfigures. The Coulomb interaction is taken into account.}
\end{figure}

 
The EPI, however, yields results of 0.05 for the resonance at 0.55 eV and 0.02 for the resonance at 1.12 eV when Coulomb interaction is turned off and gives only slightly higher values of 0.13 and 0.23 for the respective resonances when the Coulomb interaction is included, \text{e.g.}, for the $E^d_{\alpha^\prime\longrightarrow\delta^{\prime}}$ and $E^d_{\alpha\longrightarrow\alpha^{\prime}}$ resonances. Consequently, they are both classified predominantly single-particle-like. For comparison, in Fig.~\ref{fig:real_space_d2}c and Fig.~\ref{fig:real_space_d2}f we present real space induced-charge patterns for the 2.98 eV and 3.06 eV resonances, which are typical single-particle-like excitations (which barely change their spectral position both with doping and with Coulomb interaction scaling).
The distributions of induced charge for the resonances $P_1$ and $P_2$ in the same nanoantenna doped with 20 electrons also show regular patterns (see Fig.~\ref{fig:real_space_d20}). These resonances, however, exhibit much higher EPIs of 0.49 and 0.78, respectively, which is in accordance with the qualitative discussion of Fig.~\ref{fig:max_abs}. The EPI takes a comparably low value for the $P_1$ resonance because a prominent single-particle-like transition is located at the same spectral position that also contributes to the excitation. From Fig.~\ref{fig:real_space_d2} we conclude, that the presence of electron-electron interaction is apparently not a necessary prerequisite for collective charge oscillation in real space. Since the existence of plasmons, however, is linked to long-range electron-electron interaction energy~\cite{Pines1952}, we prefer to define a plasmon at the nanoscale not by charge occupation characteristics in real space, but rather by coherence considerations in energy space. From the charge occupation patterns in real space only, one can hardly tell apart bare single-particle-like resonances (Fig.~\ref{fig:real_space_d2}b) from dressed single-particle-like resonances (Fig.~\ref{fig:real_space_d2}e) and plasmonic resonances (Fig.~\ref{fig:real_space_d20}a).

\subsection{Energy Space Dynamics}
\begin{figure*}
\includegraphics[width=0.9\linewidth]{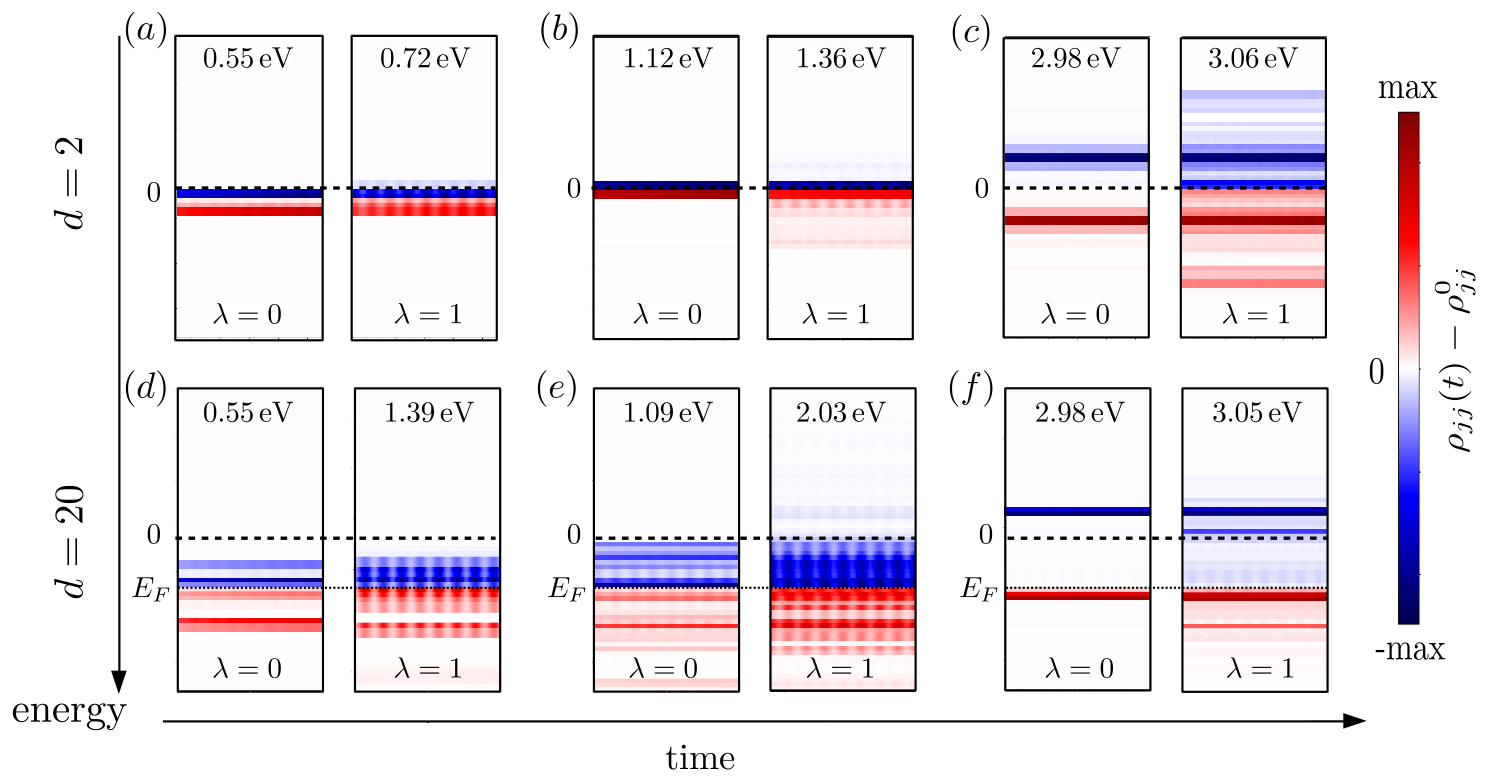}
\caption{\label{fig:energy_space_d2_d20} Population difference $\rho_{jj}(t)-\rho_{jj}^0$ of the energy states of the triangular 270-atom graphene nanoantenna doped with 2 electrons (upper row) or 20 electrons (lower row) under CW illumination for the last five optical cycles of the simulation period with respect to the ground state $\rho^0$. The illumination frequencies coincide with those in Figs~\ref{fig:real_space_d2} and~\ref{fig:real_space_d20}. The left panel in each subfigure shows results for non-interacting electrons ($\lambda=0$), whereas the right panel shows data for interacting electrons ($\lambda=1$).}
\end{figure*}

Finally, to get a systematic picture of the population dynamics in the nanoantenna under consideration, we examine the energy space dynamics as well. In Fig.~\ref{fig:energy_space_d2_d20} we depict the difference of the occupation of the eigenstates located near the Fermi energy with respect to the initial state of the system as a function of time. We show the last five optical cycles of the simulation period both for non-interacting ($\lambda=0$) and for interacting ($\lambda=1$) electrons. Without the Coulomb interaction, one can recognize pairs or groups of states among which charge transfer appears. When the Coulomb interaction is taken into account, the oscillatory movement of occupation emerges on top. This "sloshing" effect appears both in the two-fold and 20-fold doped nanoantenna, but it is more pronounced in the latter one.

The close relation between the subfigures in Fig.~\ref{fig:energy_space_d2_d20} corresponding to the case of  non-interating electrons (left) and interacting electrons (right) reveals that the investigated resonances share a significant component of single-particle-like transitions. Population transfer between the pairs of states that contribute to a transition in the non-interacting case, is present also in the interacting case. On top of that, the sloshing-type (Figs.~\ref{fig:energy_space_d2_d20}b and d) or even the inversion-type (Figs.~\ref{fig:energy_space_d2_d20}c and f) population dynamics in the interacting case may engage additional states, building up the complex collective interaction-mediated response. The EPI is a measure of such collective influence and naturally its larger values correspond to larger differences between the dynamics in the case of non-interacting and interacting electrons.

\section{\label{sec:conclusions}Conclusions and summary}
We have inspected selected resonances in a triangular graphene nanoantenna to distinguish those of predominantly single-particle-like from those of collective character. Based on the fact, that the resonances corresponding to the transitions $\alpha \longrightarrow \alpha^{\prime}$ and $\alpha^{\prime} \longrightarrow \delta^{\prime}$ are clearly visible in the spectrum of an undoped/two-fold doped structure when the Coulomb interaction is turned off, we claim that they are predominantly of a single-particle-like nature. Even though their real space charge distributions show dipolar patterns, the EPI does not classify these resonances as plasmonic, like the GPI, for instance, does. This classification is supported by the fact, that the real space charge distribution pattern is also dipolar for non-interacting electrons. Moreover, said resonances exhibit a qualitatively different dependence on the Coulomb scaling parameter $\lambda$, than the resonances in 10-/20-fold doped structures which appear to be plasmonic. Their energy patterns show a stronger "sloshing" behavior in the strongly doped structure as compared to a weak presence of this effect in the two-fold doped nanoantenna. This behavior arises due to Coulomb interactions among electrons, whose influence on their dynamics is significantly stronger in the case of collective plasmonic resonances.
  \par 
The EPI takes a low value for $\alpha \longrightarrow \alpha^{\prime}$ and $\alpha^{\prime} \longrightarrow \delta^{\prime}$ resonances and significantly higher values for the $P_1$ and $P_2$ resonances, which is in agreement with the predictions that those pairs of resonances are of different nature. This is a qualitative difference beyond analyzing the real space induced charge patterns that requires a careful look at both the absorption spectra of the nanostructure and the coherence dynamics in energy basis. \par 
These conclusions ultimately raise the question, whether one can find plasmons in the sense of the EPI, \textit{i.e.}, originating from long-range electron-electron interactions, in weakly or moderately doped nanoantennas at all. Our research rather suggests that in these structures, resonances are predominantly single-particle-like and only moderatly dressed by interaction effects. To obtain truly plasmonic resonances in the sense of the EPI, which in the first place emerge due to long-range electron-electron interaction, one needs heavy doping.

\begin{acknowledgments}
M.M.M. acknowledges financial support through the Research
Travel Grant by the Karlsruhe House of Young Scientists
(KHYS). M.M.M. and C.R. acknowledge support by the Deutsche Forschungsgemeinschaft (DFG, German Research
Foundation) project number 378579271 within project
RO 3640/8-1 and from the VolkswagenStiftung. M.M.M. is
grateful for the support of the Toru\'{n} Astrophysics/Physics
Summer Program TAPS 2019 and the PROM project no. PPI/
PRO/2018/1/00016/U/001 by the Polish National Agency
for Academic Exchange. M.M.M., M.K., and K.S. acknowledge
the hospitality of the Donostia International Physics Center.
M.K. acknowledges the financial support from the National Science Centre, Poland (grant no. 2016/23/D/ST2/02064). M.P. and A.A. acknowledge
financial support by project PID2019-105488GB-I00 of the
Spanish Ministry of Science and Innovation, and the Gobierno
Vasco UPV/EHU project IT1246-19. K.S. acknowledges
support from the National Science Centre, Poland (project no.
2016/23/G/ST3/0404).
\end{acknowledgments}

\section*{Data Availability}
The data that support the findings of this study are available from the corresponding author upon reasonable request.



\section*{References}
\bibliography{main}

\providecommand{\noopsort}[1]{}\providecommand{\singleletter}[1]{#1}%
\begin{thebibliography}{51}%
\makeatletter
\providecommand \@ifxundefined [1]{%
 \@ifx{#1\undefined}
}%
\providecommand \@ifnum [1]{%
 \ifnum #1\expandafter \@firstoftwo
 \else \expandafter \@secondoftwo
 \fi
}%
\providecommand \@ifx [1]{%
 \ifx #1\expandafter \@firstoftwo
 \else \expandafter \@secondoftwo
 \fi
}%
\providecommand \natexlab [1]{#1}%
\providecommand \enquote  [1]{``#1''}%
\providecommand \bibnamefont  [1]{#1}%
\providecommand \bibfnamefont [1]{#1}%
\providecommand \citenamefont [1]{#1}%
\providecommand \href@noop [0]{\@secondoftwo}%
\providecommand \href [0]{\begingroup \@sanitize@url \@href}%
\providecommand \@href[1]{\@@startlink{#1}\@@href}%
\providecommand \@@href[1]{\endgroup#1\@@endlink}%
\providecommand \@sanitize@url [0]{\catcode `\\12\catcode `\$12\catcode
  `\&12\catcode `\#12\catcode `\^12\catcode `\_12\catcode `\%12\relax}%
\providecommand \@@startlink[1]{}%
\providecommand \@@endlink[0]{}%
\providecommand \url  [0]{\begingroup\@sanitize@url \@url }%
\providecommand \@url [1]{\endgroup\@href {#1}{\urlprefix }}%
\providecommand \urlprefix  [0]{URL }%
\providecommand \Eprint [0]{\href }%
\providecommand \doibase [0]{http://dx.doi.org/}%
\providecommand \selectlanguage [0]{\@gobble}%
\providecommand \bibinfo  [0]{\@secondoftwo}%
\providecommand \bibfield  [0]{\@secondoftwo}%
\providecommand \translation [1]{[#1]}%
\providecommand \BibitemOpen [0]{}%
\providecommand \bibitemStop [0]{}%
\providecommand \bibitemNoStop [0]{.\EOS\space}%
\providecommand \EOS [0]{\spacefactor3000\relax}%
\providecommand \BibitemShut  [1]{\csname bibitem#1\endcsname}%
\let\auto@bib@innerbib\@empty
\bibitem [{\citenamefont {M\"uller}\ \emph {et~al.}(2020)\citenamefont
  {M\"uller}, \citenamefont {Kosik}, \citenamefont {Pelc}, \citenamefont
  {Bryant}, \citenamefont {Ayuela}, \citenamefont {Rockstuhl},\ and\
  \citenamefont {S{\l}owik}}]{Mueller2020}%
  \BibitemOpen
  \bibfield  {author} {\bibinfo {author} {\bibfnamefont {M.~M.}\ \bibnamefont
  {M\"uller}}, \bibinfo {author} {\bibfnamefont {M.}~\bibnamefont {Kosik}},
  \bibinfo {author} {\bibfnamefont {M.}~\bibnamefont {Pelc}}, \bibinfo {author}
  {\bibfnamefont {G.~W.}\ \bibnamefont {Bryant}}, \bibinfo {author}
  {\bibfnamefont {A.}~\bibnamefont {Ayuela}}, \bibinfo {author} {\bibfnamefont
  {C.}~\bibnamefont {Rockstuhl}}, \ and\ \bibinfo {author} {\bibfnamefont
  {K.}~\bibnamefont {S{\l}owik}},\ }\bibfield  {title} {\enquote {\bibinfo
  {title} {Energy-based plasmonicity index to characterize optical resonances
  in nanostructures},}\ }\href@noop {} {\bibfield  {journal} {\bibinfo
  {journal} {J. Phys. Chem. C}\ }\textbf {\bibinfo {volume} {124}},\ \bibinfo
  {pages} {24331--24343} (\bibinfo {year} {2020})}\BibitemShut {NoStop}%
\bibitem [{\citenamefont {Cazalilla}\ \emph {et~al.}(2000)\citenamefont
  {Cazalilla}, \citenamefont {Dolado}, \citenamefont {Rubio},\ and\
  \citenamefont {Echenique}}]{Cazalilla2000}%
  \BibitemOpen
  \bibfield  {author} {\bibinfo {author} {\bibfnamefont {M.~A.}\ \bibnamefont
  {Cazalilla}}, \bibinfo {author} {\bibfnamefont {J.~S.}\ \bibnamefont
  {Dolado}}, \bibinfo {author} {\bibfnamefont {A.}~\bibnamefont {Rubio}}, \
  and\ \bibinfo {author} {\bibfnamefont {P.~M.}\ \bibnamefont {Echenique}},\
  }\bibfield  {title} {\enquote {\bibinfo {title} {Plasmonic excitations in
  noble metals: The case of ag},}\ }\href {\doibase 10.1103/PhysRevB.61.8033}
  {\bibfield  {journal} {\bibinfo  {journal} {Phys. Rev. B}\ }\textbf {\bibinfo
  {volume} {61}},\ \bibinfo {pages} {8033--8042} (\bibinfo {year}
  {2000})}\BibitemShut {NoStop}%
\bibitem [{\citenamefont {Maier}(2007)}]{Maier2007}%
  \BibitemOpen
  \bibfield  {author} {\bibinfo {author} {\bibfnamefont {S.~A.}\ \bibnamefont
  {Maier}},\ }\href@noop {} {\emph {\bibinfo {title} {Plasmonics: fundamentals
  and applications}}}\ (\bibinfo  {publisher} {Springer Science \& Business
  Media},\ \bibinfo {year} {2007})\BibitemShut {NoStop}%
\bibitem [{\citenamefont {Pelton}, \citenamefont {Aizpurua},\ and\
  \citenamefont {Bryant}(2008)}]{Pelton2008}%
  \BibitemOpen
  \bibfield  {author} {\bibinfo {author} {\bibfnamefont {M.}~\bibnamefont
  {Pelton}}, \bibinfo {author} {\bibfnamefont {J.}~\bibnamefont {Aizpurua}}, \
  and\ \bibinfo {author} {\bibfnamefont {G.}~\bibnamefont {Bryant}},\
  }\bibfield  {title} {\enquote {\bibinfo {title} {Metal-nanoparticle
  plasmonics},}\ }\href@noop {} {\bibfield  {journal} {\bibinfo  {journal}
  {Laser \& Phot. Rev.}\ }\textbf {\bibinfo {volume} {2}},\ \bibinfo {pages}
  {136--159} (\bibinfo {year} {2008})}\BibitemShut {NoStop}%
\bibitem [{\citenamefont {Giannini}\ \emph {et~al.}(2011)\citenamefont
  {Giannini}, \citenamefont {Fern{\'a}ndez-Dom{\'\i}nguez}, \citenamefont
  {Heck},\ and\ \citenamefont {Maier}}]{Giannini2011}%
  \BibitemOpen
  \bibfield  {author} {\bibinfo {author} {\bibfnamefont {V.}~\bibnamefont
  {Giannini}}, \bibinfo {author} {\bibfnamefont {A.~I.}\ \bibnamefont
  {Fern{\'a}ndez-Dom{\'\i}nguez}}, \bibinfo {author} {\bibfnamefont {S.~C.}\
  \bibnamefont {Heck}}, \ and\ \bibinfo {author} {\bibfnamefont {S.~A.}\
  \bibnamefont {Maier}},\ }\bibfield  {title} {\enquote {\bibinfo {title}
  {Plasmonic nanoantennas: fundamentals and their use in controlling the
  radiative properties of nanoemitters},}\ }\href@noop {} {\bibfield  {journal}
  {\bibinfo  {journal} {Chem. Rev.}\ }\textbf {\bibinfo {volume} {111}},\
  \bibinfo {pages} {3888--3912} (\bibinfo {year} {2011})}\BibitemShut {NoStop}%
\bibitem [{\citenamefont {Koppens}, \citenamefont {Chang},\ and\ \citenamefont
  {Garc{\'\i}a~de Abajo}(2011)}]{Koppens2011}%
  \BibitemOpen
  \bibfield  {author} {\bibinfo {author} {\bibfnamefont {F.~H.}\ \bibnamefont
  {Koppens}}, \bibinfo {author} {\bibfnamefont {D.~E.}\ \bibnamefont {Chang}},
  \ and\ \bibinfo {author} {\bibfnamefont {F.~J.}\ \bibnamefont {Garc{\'\i}a~de
  Abajo}},\ }\bibfield  {title} {\enquote {\bibinfo {title} {Graphene
  plasmonics: a platform for strong light--matter interactions},}\ }\href@noop
  {} {\bibfield  {journal} {\bibinfo  {journal} {Nano Lett.}\ }\textbf
  {\bibinfo {volume} {11}},\ \bibinfo {pages} {3370--3377} (\bibinfo {year}
  {2011})}\BibitemShut {NoStop}%
\bibitem [{\citenamefont {Tame}\ \emph {et~al.}(2013)\citenamefont {Tame},
  \citenamefont {McEnery}, \citenamefont {{\"O}zdemir}, \citenamefont {Lee},
  \citenamefont {Maier},\ and\ \citenamefont {Kim}}]{Tame2013}%
  \BibitemOpen
  \bibfield  {author} {\bibinfo {author} {\bibfnamefont {M.~S.}\ \bibnamefont
  {Tame}}, \bibinfo {author} {\bibfnamefont {K.}~\bibnamefont {McEnery}},
  \bibinfo {author} {\bibfnamefont {{\c{S}}.}~\bibnamefont {{\"O}zdemir}},
  \bibinfo {author} {\bibfnamefont {J.}~\bibnamefont {Lee}}, \bibinfo {author}
  {\bibfnamefont {S.~A.}\ \bibnamefont {Maier}}, \ and\ \bibinfo {author}
  {\bibfnamefont {M.}~\bibnamefont {Kim}},\ }\bibfield  {title} {\enquote
  {\bibinfo {title} {Quantum plasmonics},}\ }\href@noop {} {\bibfield
  {journal} {\bibinfo  {journal} {Nat. Phys.}\ }\textbf {\bibinfo {volume}
  {9}},\ \bibinfo {pages} {329--340} (\bibinfo {year} {2013})}\BibitemShut
  {NoStop}%
\bibitem [{\citenamefont {Bryant}, \citenamefont {Waks},\ and\ \citenamefont
  {Krenn}(2014)}]{Bryant2014}%
  \BibitemOpen
  \bibfield  {author} {\bibinfo {author} {\bibfnamefont {G.~W.}\ \bibnamefont
  {Bryant}}, \bibinfo {author} {\bibfnamefont {E.}~\bibnamefont {Waks}}, \ and\
  \bibinfo {author} {\bibfnamefont {J.~R.}\ \bibnamefont {Krenn}},\ }\bibfield
  {title} {\enquote {\bibinfo {title} {Plasmonics: The rise of quantum
  effects},}\ }\href@noop {} {\bibfield  {journal} {\bibinfo  {journal} {Optics
  and Photonics News}\ }\textbf {\bibinfo {volume} {25}},\ \bibinfo {pages}
  {50--53} (\bibinfo {year} {2014})}\BibitemShut {NoStop}%
\bibitem [{\citenamefont {Bozhevolnyi}\ and\ \citenamefont
  {Mortensen}(2017)}]{Bolzhevolnyi2017}%
  \BibitemOpen
  \bibfield  {author} {\bibinfo {author} {\bibfnamefont {S.~I.}\ \bibnamefont
  {Bozhevolnyi}}\ and\ \bibinfo {author} {\bibfnamefont {N.~A.}\ \bibnamefont
  {Mortensen}},\ }\bibfield  {title} {\enquote {\bibinfo {title} {Plasmonics
  for emerging quantum technologies},}\ }\href@noop {} {\bibfield  {journal}
  {\bibinfo  {journal} {Nanophot.}\ }\textbf {\bibinfo {volume} {6}},\ \bibinfo
  {pages} {1185--1188} (\bibinfo {year} {2017})}\BibitemShut {NoStop}%
\bibitem [{\citenamefont {Takahara}\ \emph {et~al.}(1997)\citenamefont
  {Takahara}, \citenamefont {Yamagishi}, \citenamefont {Taki}, \citenamefont
  {Morimoto},\ and\ \citenamefont {Kobayashi}}]{Takahara1997}%
  \BibitemOpen
  \bibfield  {author} {\bibinfo {author} {\bibfnamefont {J.}~\bibnamefont
  {Takahara}}, \bibinfo {author} {\bibfnamefont {S.}~\bibnamefont {Yamagishi}},
  \bibinfo {author} {\bibfnamefont {H.}~\bibnamefont {Taki}}, \bibinfo {author}
  {\bibfnamefont {A.}~\bibnamefont {Morimoto}}, \ and\ \bibinfo {author}
  {\bibfnamefont {T.}~\bibnamefont {Kobayashi}},\ }\bibfield  {title} {\enquote
  {\bibinfo {title} {Guiding of a one-dimensional optical beam with nanometer
  diameter},}\ }\href@noop {} {\bibfield  {journal} {\bibinfo  {journal} {Opt.
  Lett.}\ }\textbf {\bibinfo {volume} {22}},\ \bibinfo {pages} {475--477}
  (\bibinfo {year} {1997})}\BibitemShut {NoStop}%
\bibitem [{\citenamefont {Gramotnev}\ and\ \citenamefont
  {Bozhevolnyi}(2010)}]{Gramotnev2010}%
  \BibitemOpen
  \bibfield  {author} {\bibinfo {author} {\bibfnamefont {D.~K.}\ \bibnamefont
  {Gramotnev}}\ and\ \bibinfo {author} {\bibfnamefont {S.~I.}\ \bibnamefont
  {Bozhevolnyi}},\ }\bibfield  {title} {\enquote {\bibinfo {title} {Plasmonics
  beyond the diffraction limit},}\ }\href@noop {} {\bibfield  {journal}
  {\bibinfo  {journal} {Nat. Phot.}\ }\textbf {\bibinfo {volume} {4}},\
  \bibinfo {pages} {83} (\bibinfo {year} {2010})}\BibitemShut {NoStop}%
\bibitem [{\citenamefont {Novotny}\ and\ \citenamefont
  {Hecht}(2012)}]{Novotny2012}%
  \BibitemOpen
  \bibfield  {author} {\bibinfo {author} {\bibfnamefont {L.}~\bibnamefont
  {Novotny}}\ and\ \bibinfo {author} {\bibfnamefont {B.}~\bibnamefont
  {Hecht}},\ }\href@noop {} {\emph {\bibinfo {title} {Principles of
  Nano-Optics}}}\ (\bibinfo  {publisher} {Cambridge university press},\
  \bibinfo {year} {2012})\BibitemShut {NoStop}%
\bibitem [{\citenamefont {Melendez}\ \emph {et~al.}(1996)\citenamefont
  {Melendez}, \citenamefont {Carr}, \citenamefont {Bartholomew}, \citenamefont
  {Kukanskis}, \citenamefont {Elkind}, \citenamefont {Yee}, \citenamefont
  {Furlong},\ and\ \citenamefont {Woodbury}}]{Melendez1996}%
  \BibitemOpen
  \bibfield  {author} {\bibinfo {author} {\bibfnamefont {J.}~\bibnamefont
  {Melendez}}, \bibinfo {author} {\bibfnamefont {R.}~\bibnamefont {Carr}},
  \bibinfo {author} {\bibfnamefont {D.~U.}\ \bibnamefont {Bartholomew}},
  \bibinfo {author} {\bibfnamefont {K.}~\bibnamefont {Kukanskis}}, \bibinfo
  {author} {\bibfnamefont {J.}~\bibnamefont {Elkind}}, \bibinfo {author}
  {\bibfnamefont {S.}~\bibnamefont {Yee}}, \bibinfo {author} {\bibfnamefont
  {C.}~\bibnamefont {Furlong}}, \ and\ \bibinfo {author} {\bibfnamefont
  {R.}~\bibnamefont {Woodbury}},\ }\bibfield  {title} {\enquote {\bibinfo
  {title} {A commercial solution for surface plasmon sensing},}\ }\href@noop {}
  {\bibfield  {journal} {\bibinfo  {journal} {Sens. Actuator B Chem.}\ }\textbf
  {\bibinfo {volume} {35}},\ \bibinfo {pages} {212--216} (\bibinfo {year}
  {1996})}\BibitemShut {NoStop}%
\bibitem [{\citenamefont {Awazu}\ \emph {et~al.}(2007)\citenamefont {Awazu},
  \citenamefont {Rockstuhl}, \citenamefont {Fujimaki}, \citenamefont {Fukuda},
  \citenamefont {Tominaga}, \citenamefont {Komatsubara}, \citenamefont
  {Ikeda},\ and\ \citenamefont {Ohki}}]{Awazu2007}%
  \BibitemOpen
  \bibfield  {author} {\bibinfo {author} {\bibfnamefont {K.}~\bibnamefont
  {Awazu}}, \bibinfo {author} {\bibfnamefont {C.}~\bibnamefont {Rockstuhl}},
  \bibinfo {author} {\bibfnamefont {M.}~\bibnamefont {Fujimaki}}, \bibinfo
  {author} {\bibfnamefont {N.}~\bibnamefont {Fukuda}}, \bibinfo {author}
  {\bibfnamefont {J.}~\bibnamefont {Tominaga}}, \bibinfo {author}
  {\bibfnamefont {T.}~\bibnamefont {Komatsubara}}, \bibinfo {author}
  {\bibfnamefont {T.}~\bibnamefont {Ikeda}}, \ and\ \bibinfo {author}
  {\bibfnamefont {Y.}~\bibnamefont {Ohki}},\ }\bibfield  {title} {\enquote
  {\bibinfo {title} {High sensitivity sensors made of perforated waveguides},}\
  }\href@noop {} {\bibfield  {journal} {\bibinfo  {journal} {Opt. Expr.}\
  }\textbf {\bibinfo {volume} {15}},\ \bibinfo {pages} {2592--2597} (\bibinfo
  {year} {2007})}\BibitemShut {NoStop}%
\bibitem [{\citenamefont {Lee}\ \emph {et~al.}(2016)\citenamefont {Lee},
  \citenamefont {Dieleman}, \citenamefont {Lee}, \citenamefont {Rockstuhl},
  \citenamefont {Maier},\ and\ \citenamefont {Tame}}]{Lee2016}%
  \BibitemOpen
  \bibfield  {author} {\bibinfo {author} {\bibfnamefont {C.}~\bibnamefont
  {Lee}}, \bibinfo {author} {\bibfnamefont {F.}~\bibnamefont {Dieleman}},
  \bibinfo {author} {\bibfnamefont {J.}~\bibnamefont {Lee}}, \bibinfo {author}
  {\bibfnamefont {C.}~\bibnamefont {Rockstuhl}}, \bibinfo {author}
  {\bibfnamefont {S.~A.}\ \bibnamefont {Maier}}, \ and\ \bibinfo {author}
  {\bibfnamefont {M.}~\bibnamefont {Tame}},\ }\bibfield  {title} {\enquote
  {\bibinfo {title} {Quantum plasmonic sensing: beyond the shot-noise and
  diffraction limit},}\ }\href@noop {} {\bibfield  {journal} {\bibinfo
  {journal} {ACS Photonics}\ }\textbf {\bibinfo {volume} {3}},\ \bibinfo
  {pages} {992--999} (\bibinfo {year} {2016})}\BibitemShut {NoStop}%
\bibitem [{\citenamefont {Kim}\ \emph {et~al.}(2014)\citenamefont {Kim},
  \citenamefont {Yu}, \citenamefont {Choi},\ and\ \citenamefont
  {Choi}}]{Kim2014}%
  \BibitemOpen
  \bibfield  {author} {\bibinfo {author} {\bibfnamefont {J.~T.}\ \bibnamefont
  {Kim}}, \bibinfo {author} {\bibfnamefont {Y.-J.}\ \bibnamefont {Yu}},
  \bibinfo {author} {\bibfnamefont {H.}~\bibnamefont {Choi}}, \ and\ \bibinfo
  {author} {\bibfnamefont {C.-G.}\ \bibnamefont {Choi}},\ }\bibfield  {title}
  {\enquote {\bibinfo {title} {Graphene-based plasmonic photodetector for
  photonic integrated circuits},}\ }\href@noop {} {\bibfield  {journal}
  {\bibinfo  {journal} {Opt. Expr.}\ }\textbf {\bibinfo {volume} {22}},\
  \bibinfo {pages} {803--808} (\bibinfo {year} {2014})}\BibitemShut {NoStop}%
\bibitem [{\citenamefont {Brongersma}, \citenamefont {Halas},\ and\
  \citenamefont {Nordlander}(2015)}]{Brongersma2015}%
  \BibitemOpen
  \bibfield  {author} {\bibinfo {author} {\bibfnamefont {M.~L.}\ \bibnamefont
  {Brongersma}}, \bibinfo {author} {\bibfnamefont {N.~J.}\ \bibnamefont
  {Halas}}, \ and\ \bibinfo {author} {\bibfnamefont {P.}~\bibnamefont
  {Nordlander}},\ }\bibfield  {title} {\enquote {\bibinfo {title}
  {Plasmon-induced hot carrier science and technology},}\ }\href@noop {}
  {\bibfield  {journal} {\bibinfo  {journal} {Nat. Nanotechn.}\ }\textbf
  {\bibinfo {volume} {10}},\ \bibinfo {pages} {25} (\bibinfo {year}
  {2015})}\BibitemShut {NoStop}%
\bibitem [{\citenamefont {Zhang}\ \emph {et~al.}(2015)\citenamefont {Zhang},
  \citenamefont {Zhu}, \citenamefont {Liu}, \citenamefont {Yuan},\ and\
  \citenamefont {Qin}}]{Zhang2015}%
  \BibitemOpen
  \bibfield  {author} {\bibinfo {author} {\bibfnamefont {J.}~\bibnamefont
  {Zhang}}, \bibinfo {author} {\bibfnamefont {Z.}~\bibnamefont {Zhu}}, \bibinfo
  {author} {\bibfnamefont {W.}~\bibnamefont {Liu}}, \bibinfo {author}
  {\bibfnamefont {X.}~\bibnamefont {Yuan}}, \ and\ \bibinfo {author}
  {\bibfnamefont {S.}~\bibnamefont {Qin}},\ }\bibfield  {title} {\enquote
  {\bibinfo {title} {Towards photodetection with high efficiency and tunable
  spectral selectivity: graphene plasmonics for light trapping and absorption
  engineering},}\ }\href {\doibase 10.1039/C5NR03060A} {\bibfield  {journal}
  {\bibinfo  {journal} {Nanoscale}\ }\textbf {\bibinfo {volume} {7}},\ \bibinfo
  {pages} {13530--13536} (\bibinfo {year} {2015})}\BibitemShut {NoStop}%
\bibitem [{\citenamefont {Yu}\ \emph {et~al.}(2016)\citenamefont {Yu},
  \citenamefont {Wu}, \citenamefont {Ashalley}, \citenamefont {Govorov},\ and\
  \citenamefont {Wang}}]{Yu2016}%
  \BibitemOpen
  \bibfield  {author} {\bibinfo {author} {\bibfnamefont {P.}~\bibnamefont
  {Yu}}, \bibinfo {author} {\bibfnamefont {J.}~\bibnamefont {Wu}}, \bibinfo
  {author} {\bibfnamefont {E.}~\bibnamefont {Ashalley}}, \bibinfo {author}
  {\bibfnamefont {A.}~\bibnamefont {Govorov}}, \ and\ \bibinfo {author}
  {\bibfnamefont {Z.}~\bibnamefont {Wang}},\ }\bibfield  {title} {\enquote
  {\bibinfo {title} {Dual-band absorber for multispectral plasmon-enhanced
  infrared photodetection},}\ }\href@noop {} {\bibfield  {journal} {\bibinfo
  {journal} {J. Phys. D: Appl. Phys.}\ }\textbf {\bibinfo {volume} {49}},\
  \bibinfo {pages} {365101} (\bibinfo {year} {2016})}\BibitemShut {NoStop}%
\bibitem [{\citenamefont {Hirsch}\ \emph {et~al.}(2003)\citenamefont {Hirsch},
  \citenamefont {Stafford}, \citenamefont {Bankson}, \citenamefont {Sershen},
  \citenamefont {Rivera}, \citenamefont {Price}, \citenamefont {Hazle},
  \citenamefont {Halas},\ and\ \citenamefont {West}}]{Hirsch2003}%
  \BibitemOpen
  \bibfield  {author} {\bibinfo {author} {\bibfnamefont {L.~R.}\ \bibnamefont
  {Hirsch}}, \bibinfo {author} {\bibfnamefont {R.~J.}\ \bibnamefont
  {Stafford}}, \bibinfo {author} {\bibfnamefont {J.~A.}\ \bibnamefont
  {Bankson}}, \bibinfo {author} {\bibfnamefont {S.~R.}\ \bibnamefont
  {Sershen}}, \bibinfo {author} {\bibfnamefont {B.}~\bibnamefont {Rivera}},
  \bibinfo {author} {\bibfnamefont {R.}~\bibnamefont {Price}}, \bibinfo
  {author} {\bibfnamefont {J.~D.}\ \bibnamefont {Hazle}}, \bibinfo {author}
  {\bibfnamefont {N.~J.}\ \bibnamefont {Halas}}, \ and\ \bibinfo {author}
  {\bibfnamefont {J.~L.}\ \bibnamefont {West}},\ }\bibfield  {title} {\enquote
  {\bibinfo {title} {Nanoshell-mediated near-infrared thermal therapy of tumors
  under magnetic resonance guidance},}\ }\href@noop {} {\bibfield  {journal}
  {\bibinfo  {journal} {Proceedings of the National Academy of Sciences}\
  }\textbf {\bibinfo {volume} {100}},\ \bibinfo {pages} {13549--13554}
  (\bibinfo {year} {2003})}\BibitemShut {NoStop}%
\bibitem [{\citenamefont {Qian}\ \emph {et~al.}(2008)\citenamefont {Qian},
  \citenamefont {Peng}, \citenamefont {Ansari}, \citenamefont {Yin-Goen},
  \citenamefont {Chen}, \citenamefont {Shin}, \citenamefont {Yang},
  \citenamefont {Young}, \citenamefont {Wang},\ and\ \citenamefont
  {Nie}}]{Qian2008}%
  \BibitemOpen
  \bibfield  {author} {\bibinfo {author} {\bibfnamefont {X.}~\bibnamefont
  {Qian}}, \bibinfo {author} {\bibfnamefont {X.-H.}\ \bibnamefont {Peng}},
  \bibinfo {author} {\bibfnamefont {D.~O.}\ \bibnamefont {Ansari}}, \bibinfo
  {author} {\bibfnamefont {Q.}~\bibnamefont {Yin-Goen}}, \bibinfo {author}
  {\bibfnamefont {G.~Z.}\ \bibnamefont {Chen}}, \bibinfo {author}
  {\bibfnamefont {D.~M.}\ \bibnamefont {Shin}}, \bibinfo {author}
  {\bibfnamefont {L.}~\bibnamefont {Yang}}, \bibinfo {author} {\bibfnamefont
  {A.~N.}\ \bibnamefont {Young}}, \bibinfo {author} {\bibfnamefont {M.~D.}\
  \bibnamefont {Wang}}, \ and\ \bibinfo {author} {\bibfnamefont
  {S.}~\bibnamefont {Nie}},\ }\bibfield  {title} {\enquote {\bibinfo {title}
  {In vivo tumor targeting and spectroscopic detection with surface-enhanced
  raman nanoparticle tags},}\ }\href@noop {} {\bibfield  {journal} {\bibinfo
  {journal} {Nat. Biotechn.}\ }\textbf {\bibinfo {volume} {26}},\ \bibinfo
  {pages} {83--90} (\bibinfo {year} {2008})}\BibitemShut {NoStop}%
\bibitem [{\citenamefont {Ju}\ \emph {et~al.}(2011)\citenamefont {Ju},
  \citenamefont {Geng}, \citenamefont {Horng}, \citenamefont {Girit},
  \citenamefont {Martin}, \citenamefont {Hao}, \citenamefont {Bechtel},
  \citenamefont {Liang}, \citenamefont {Zettl}, \citenamefont {Shen} \emph
  {et~al.}}]{Ju2011}%
  \BibitemOpen
  \bibfield  {author} {\bibinfo {author} {\bibfnamefont {L.}~\bibnamefont
  {Ju}}, \bibinfo {author} {\bibfnamefont {B.}~\bibnamefont {Geng}}, \bibinfo
  {author} {\bibfnamefont {J.}~\bibnamefont {Horng}}, \bibinfo {author}
  {\bibfnamefont {C.}~\bibnamefont {Girit}}, \bibinfo {author} {\bibfnamefont
  {M.}~\bibnamefont {Martin}}, \bibinfo {author} {\bibfnamefont
  {Z.}~\bibnamefont {Hao}}, \bibinfo {author} {\bibfnamefont {H.~A.}\
  \bibnamefont {Bechtel}}, \bibinfo {author} {\bibfnamefont {X.}~\bibnamefont
  {Liang}}, \bibinfo {author} {\bibfnamefont {A.}~\bibnamefont {Zettl}},
  \bibinfo {author} {\bibfnamefont {Y.~R.}\ \bibnamefont {Shen}},  \emph
  {et~al.},\ }\bibfield  {title} {\enquote {\bibinfo {title} {Graphene
  plasmonics for tunable terahertz metamaterials},}\ }\href@noop {} {\bibfield
  {journal} {\bibinfo  {journal} {Nat. Nanotech.}\ }\textbf {\bibinfo {volume}
  {6}},\ \bibinfo {pages} {630--634} (\bibinfo {year} {2011})}\BibitemShut
  {NoStop}%
\bibitem [{\citenamefont {Hess}\ \emph {et~al.}(2012)\citenamefont {Hess},
  \citenamefont {Pendry}, \citenamefont {Maier}, \citenamefont {Oulton},
  \citenamefont {Hamm},\ and\ \citenamefont {Tsakmakidis}}]{Hess2012}%
  \BibitemOpen
  \bibfield  {author} {\bibinfo {author} {\bibfnamefont {O.}~\bibnamefont
  {Hess}}, \bibinfo {author} {\bibfnamefont {J.~B.}\ \bibnamefont {Pendry}},
  \bibinfo {author} {\bibfnamefont {S.~A.}\ \bibnamefont {Maier}}, \bibinfo
  {author} {\bibfnamefont {R.~F.}\ \bibnamefont {Oulton}}, \bibinfo {author}
  {\bibfnamefont {J.~M.}\ \bibnamefont {Hamm}}, \ and\ \bibinfo {author}
  {\bibfnamefont {K.~L.}\ \bibnamefont {Tsakmakidis}},\ }\bibfield  {title}
  {\enquote {\bibinfo {title} {Active nanoplasmonic metamaterials},}\
  }\href@noop {} {\bibfield  {journal} {\bibinfo  {journal} {Nat. Mat.}\
  }\textbf {\bibinfo {volume} {11}},\ \bibinfo {pages} {573--584} (\bibinfo
  {year} {2012})}\BibitemShut {NoStop}%
\bibitem [{\citenamefont {M{\"u}hlig}\ \emph {et~al.}(2013)\citenamefont
  {M{\"u}hlig}, \citenamefont {Cunningham}, \citenamefont {Dintinger},
  \citenamefont {Scharf}, \citenamefont {B{\"u}rgi}, \citenamefont {Lederer},\
  and\ \citenamefont {Rockstuhl}}]{Muhlig2013}%
  \BibitemOpen
  \bibfield  {author} {\bibinfo {author} {\bibfnamefont {S.}~\bibnamefont
  {M{\"u}hlig}}, \bibinfo {author} {\bibfnamefont {A.}~\bibnamefont
  {Cunningham}}, \bibinfo {author} {\bibfnamefont {J.}~\bibnamefont
  {Dintinger}}, \bibinfo {author} {\bibfnamefont {T.}~\bibnamefont {Scharf}},
  \bibinfo {author} {\bibfnamefont {T.}~\bibnamefont {B{\"u}rgi}}, \bibinfo
  {author} {\bibfnamefont {F.}~\bibnamefont {Lederer}}, \ and\ \bibinfo
  {author} {\bibfnamefont {C.}~\bibnamefont {Rockstuhl}},\ }\bibfield  {title}
  {\enquote {\bibinfo {title} {Self-assembled plasmonic metamaterials},}\
  }\href@noop {} {\bibfield  {journal} {\bibinfo  {journal} {Nanophot.}\
  }\textbf {\bibinfo {volume} {2}},\ \bibinfo {pages} {211--240} (\bibinfo
  {year} {2013})}\BibitemShut {NoStop}%
\bibitem [{\citenamefont {Koenderink}(2009)}]{Koenderink2009}%
  \BibitemOpen
  \bibfield  {author} {\bibinfo {author} {\bibfnamefont {A.~F.}\ \bibnamefont
  {Koenderink}},\ }\bibfield  {title} {\enquote {\bibinfo {title} {Plasmon
  nanoparticle array waveguides for single photon and single plasmon
  sources},}\ }\href@noop {} {\bibfield  {journal} {\bibinfo  {journal} {Nano
  Lett.}\ }\textbf {\bibinfo {volume} {9}},\ \bibinfo {pages} {4228--4233}
  (\bibinfo {year} {2009})}\BibitemShut {NoStop}%
\bibitem [{\citenamefont {Chen}, \citenamefont {Lodahl},\ and\ \citenamefont
  {Koenderink}(2010)}]{Chen2010}%
  \BibitemOpen
  \bibfield  {author} {\bibinfo {author} {\bibfnamefont {Y.}~\bibnamefont
  {Chen}}, \bibinfo {author} {\bibfnamefont {P.}~\bibnamefont {Lodahl}}, \ and\
  \bibinfo {author} {\bibfnamefont {A.~F.}\ \bibnamefont {Koenderink}},\
  }\bibfield  {title} {\enquote {\bibinfo {title} {Dynamically reconfigurable
  directionality of plasmon-based single photon sources},}\ }\href@noop {}
  {\bibfield  {journal} {\bibinfo  {journal} {Phys. Rev. B}\ }\textbf {\bibinfo
  {volume} {82}},\ \bibinfo {pages} {081402} (\bibinfo {year}
  {2010})}\BibitemShut {NoStop}%
\bibitem [{\citenamefont {García~de Abajo}(2014)}]{deAbajo2014}%
  \BibitemOpen
  \bibfield  {author} {\bibinfo {author} {\bibfnamefont {F.~J.}\ \bibnamefont
  {García~de Abajo}},\ }\bibfield  {title} {\enquote {\bibinfo {title}
  {Graphene plasmonics: Challenges and opportunities},}\ }\href {\doibase
  10.1021/ph400147y} {\bibfield  {journal} {\bibinfo  {journal} {ACS
  Photonics}\ }\textbf {\bibinfo {volume} {1}},\ \bibinfo {pages} {135--152}
  (\bibinfo {year} {2014})}\BibitemShut {NoStop}%
\bibitem [{\citenamefont {Gonçalves}\ and\ \citenamefont
  {Peres}(2016)}]{Goncalves2016}%
  \BibitemOpen
  \bibfield  {author} {\bibinfo {author} {\bibfnamefont {P.~A.~D.}\
  \bibnamefont {Gonçalves}}\ and\ \bibinfo {author} {\bibfnamefont {N.~M.~R.}\
  \bibnamefont {Peres}},\ }\href {\doibase 10.1142/9948} {\emph {\bibinfo
  {title} {An Introduction to Graphene Plasmonics}}}\ (\bibinfo  {publisher}
  {WORLD SCIENTIFIC},\ \bibinfo {year} {2016})\ \Eprint
  {http://arxiv.org/abs/https://www.worldscientific.com/doi/pdf/10.1142/9948}
  {https://www.worldscientific.com/doi/pdf/10.1142/9948} \BibitemShut {NoStop}%
\bibitem [{\citenamefont {Yamada}\ \emph {et~al.}(2013)\citenamefont {Yamada},
  \citenamefont {Kim}, \citenamefont {Ishihara},\ and\ \citenamefont
  {Hasegawa}}]{Yamada2013}%
  \BibitemOpen
  \bibfield  {author} {\bibinfo {author} {\bibfnamefont {T.}~\bibnamefont
  {Yamada}}, \bibinfo {author} {\bibfnamefont {J.}~\bibnamefont {Kim}},
  \bibinfo {author} {\bibfnamefont {M.}~\bibnamefont {Ishihara}}, \ and\
  \bibinfo {author} {\bibfnamefont {M.}~\bibnamefont {Hasegawa}},\ }\bibfield
  {title} {\enquote {\bibinfo {title} {Low-temperature graphene synthesis using
  microwave plasma cvd},}\ }\href@noop {} {\bibfield  {journal} {\bibinfo
  {journal} {J. Phys. D: Appl. Phys.}\ }\textbf {\bibinfo {volume} {46}},\
  \bibinfo {pages} {063001} (\bibinfo {year} {2013})}\BibitemShut {NoStop}%
\bibitem [{\citenamefont {Jabari}\ \emph {et~al.}(2019)\citenamefont {Jabari},
  \citenamefont {Ahmed}, \citenamefont {Liravi}, \citenamefont {Secor},
  \citenamefont {Lin},\ and\ \citenamefont {Toyserkani}}]{Jabari2019}%
  \BibitemOpen
  \bibfield  {author} {\bibinfo {author} {\bibfnamefont {E.}~\bibnamefont
  {Jabari}}, \bibinfo {author} {\bibfnamefont {F.}~\bibnamefont {Ahmed}},
  \bibinfo {author} {\bibfnamefont {F.}~\bibnamefont {Liravi}}, \bibinfo
  {author} {\bibfnamefont {E.~B.}\ \bibnamefont {Secor}}, \bibinfo {author}
  {\bibfnamefont {L.}~\bibnamefont {Lin}}, \ and\ \bibinfo {author}
  {\bibfnamefont {E.}~\bibnamefont {Toyserkani}},\ }\bibfield  {title}
  {\enquote {\bibinfo {title} {2d printing of graphene: a review},}\ }\href
  {\doibase 10.1088/2053-1583/ab29b2} {\bibfield  {journal} {\bibinfo
  {journal} {2D Materials}\ }\textbf {\bibinfo {volume} {6}},\ \bibinfo {pages}
  {042004} (\bibinfo {year} {2019})}\BibitemShut {NoStop}%
\bibitem [{\citenamefont {Thongrattanasiri}, \citenamefont {Manjavacas},\ and\
  \citenamefont {Garc{\'\i}a~de Abajo}(2012)}]{Thongrattanasiri2012}%
  \BibitemOpen
  \bibfield  {author} {\bibinfo {author} {\bibfnamefont {S.}~\bibnamefont
  {Thongrattanasiri}}, \bibinfo {author} {\bibfnamefont {A.}~\bibnamefont
  {Manjavacas}}, \ and\ \bibinfo {author} {\bibfnamefont {F.~J.}\ \bibnamefont
  {Garc{\'\i}a~de Abajo}},\ }\bibfield  {title} {\enquote {\bibinfo {title}
  {Quantum finite-size effects in graphene plasmons},}\ }\href@noop {}
  {\bibfield  {journal} {\bibinfo  {journal} {ACS Nano}\ }\textbf {\bibinfo
  {volume} {6}},\ \bibinfo {pages} {1766--1775} (\bibinfo {year}
  {2012})}\BibitemShut {NoStop}%
\bibitem [{\citenamefont {Manjavacas}, \citenamefont {Thongrattanasiri},\ and\
  \citenamefont {Garc{\'\i}a~de Abajo}(2013)}]{Manjavacas2013}%
  \BibitemOpen
  \bibfield  {author} {\bibinfo {author} {\bibfnamefont {A.}~\bibnamefont
  {Manjavacas}}, \bibinfo {author} {\bibfnamefont {S.}~\bibnamefont
  {Thongrattanasiri}}, \ and\ \bibinfo {author} {\bibfnamefont {F.~J.}\
  \bibnamefont {Garc{\'\i}a~de Abajo}},\ }\bibfield  {title} {\enquote
  {\bibinfo {title} {Plasmons driven by single electrons in graphene
  nanoislands},}\ }\href@noop {} {\bibfield  {journal} {\bibinfo  {journal}
  {Nanophot.}\ }\textbf {\bibinfo {volume} {2}},\ \bibinfo {pages} {139--151}
  (\bibinfo {year} {2013})}\BibitemShut {NoStop}%
\bibitem [{\citenamefont {Piccini}\ \emph {et~al.}(2013)\citenamefont
  {Piccini}, \citenamefont {Havenith}, \citenamefont {Broer},\ and\
  \citenamefont {Stener}}]{Piccini2013}%
  \BibitemOpen
  \bibfield  {author} {\bibinfo {author} {\bibfnamefont {G.}~\bibnamefont
  {Piccini}}, \bibinfo {author} {\bibfnamefont {R.~W.}\ \bibnamefont
  {Havenith}}, \bibinfo {author} {\bibfnamefont {R.}~\bibnamefont {Broer}}, \
  and\ \bibinfo {author} {\bibfnamefont {M.}~\bibnamefont {Stener}},\
  }\bibfield  {title} {\enquote {\bibinfo {title} {Gold nanowires: a
  time-dependent density functional assessment of plasmonic behavior},}\
  }\href@noop {} {\bibfield  {journal} {\bibinfo  {journal} {J. Phys. Chem. C}\
  }\textbf {\bibinfo {volume} {117}},\ \bibinfo {pages} {17196--17204}
  (\bibinfo {year} {2013})}\BibitemShut {NoStop}%
\bibitem [{\citenamefont {Bursi}\ \emph {et~al.}(2016)\citenamefont {Bursi},
  \citenamefont {Calzolari}, \citenamefont {Corni},\ and\ \citenamefont
  {Molinari}}]{Bursi2016}%
  \BibitemOpen
  \bibfield  {author} {\bibinfo {author} {\bibfnamefont {L.}~\bibnamefont
  {Bursi}}, \bibinfo {author} {\bibfnamefont {A.}~\bibnamefont {Calzolari}},
  \bibinfo {author} {\bibfnamefont {S.}~\bibnamefont {Corni}}, \ and\ \bibinfo
  {author} {\bibfnamefont {E.}~\bibnamefont {Molinari}},\ }\bibfield  {title}
  {\enquote {\bibinfo {title} {Quantifying the plasmonic character of optical
  excitations in nanostructures},}\ }\href@noop {} {\bibfield  {journal}
  {\bibinfo  {journal} {ACS Photonics}\ }\textbf {\bibinfo {volume} {3}},\
  \bibinfo {pages} {520--525} (\bibinfo {year} {2016})}\BibitemShut {NoStop}%
\bibitem [{\citenamefont {Noguchi}, \citenamefont {Shimamoto},\ and\
  \citenamefont {Watanabe}(2005)}]{Noguchi2005}%
  \BibitemOpen
  \bibfield  {author} {\bibinfo {author} {\bibfnamefont {T.}~\bibnamefont
  {Noguchi}}, \bibinfo {author} {\bibfnamefont {T.}~\bibnamefont {Shimamoto}},
  \ and\ \bibinfo {author} {\bibfnamefont {K.}~\bibnamefont {Watanabe}},\
  }\bibfield  {title} {\enquote {\bibinfo {title} {Photoabsorption spectra of
  graphitic nanostructures by time-dependent density-functional theory},}\
  }\href@noop {} {\bibfield  {journal} {\bibinfo  {journal} {e-Journal of
  Surface Science and Nanotechnology}\ }\textbf {\bibinfo {volume} {3}},\
  \bibinfo {pages} {439--443} (\bibinfo {year} {2005})}\BibitemShut {NoStop}%
\bibitem [{\citenamefont {Manjavacas}\ \emph {et~al.}(2013)\citenamefont
  {Manjavacas}, \citenamefont {Marchesin}, \citenamefont {Thongrattanasiri},
  \citenamefont {Koval}, \citenamefont {Nordlander}, \citenamefont
  {Sanchez-Portal},\ and\ \citenamefont {Garc{\'\i}a~de
  Abajo}}]{Manjavacas2013TDDFT}%
  \BibitemOpen
  \bibfield  {author} {\bibinfo {author} {\bibfnamefont {A.}~\bibnamefont
  {Manjavacas}}, \bibinfo {author} {\bibfnamefont {F.}~\bibnamefont
  {Marchesin}}, \bibinfo {author} {\bibfnamefont {S.}~\bibnamefont
  {Thongrattanasiri}}, \bibinfo {author} {\bibfnamefont {P.}~\bibnamefont
  {Koval}}, \bibinfo {author} {\bibfnamefont {P.}~\bibnamefont {Nordlander}},
  \bibinfo {author} {\bibfnamefont {D.}~\bibnamefont {Sanchez-Portal}}, \ and\
  \bibinfo {author} {\bibfnamefont {F.~J.}\ \bibnamefont {Garc{\'\i}a~de
  Abajo}},\ }\bibfield  {title} {\enquote {\bibinfo {title} {Tunable molecular
  plasmons in polycyclic aromatic hydrocarbons},}\ }\href@noop {} {\bibfield
  {journal} {\bibinfo  {journal} {ACS Nano}\ }\textbf {\bibinfo {volume} {7}},\
  \bibinfo {pages} {3635--3643} (\bibinfo {year} {2013})}\BibitemShut {NoStop}%
\bibitem [{\citenamefont {Ezawa}(2007)}]{Ezawa2007}%
  \BibitemOpen
  \bibfield  {author} {\bibinfo {author} {\bibfnamefont {M.}~\bibnamefont
  {Ezawa}},\ }\bibfield  {title} {\enquote {\bibinfo {title} {Metallic graphene
  nanodisks: Electronic and magnetic properties},}\ }\href@noop {} {\bibfield
  {journal} {\bibinfo  {journal} {Phys. Rev. B}\ }\textbf {\bibinfo {volume}
  {76}},\ \bibinfo {pages} {245415} (\bibinfo {year} {2007})}\BibitemShut
  {NoStop}%
\bibitem [{\citenamefont {G{\"u}{\c{c}}l{\"u}}, \citenamefont {Potasz},\ and\
  \citenamefont {Hawrylak}(2010)}]{Gucclu2010}%
  \BibitemOpen
  \bibfield  {author} {\bibinfo {author} {\bibfnamefont {A.}~\bibnamefont
  {G{\"u}{\c{c}}l{\"u}}}, \bibinfo {author} {\bibfnamefont {P.}~\bibnamefont
  {Potasz}}, \ and\ \bibinfo {author} {\bibfnamefont {P.}~\bibnamefont
  {Hawrylak}},\ }\bibfield  {title} {\enquote {\bibinfo {title} {Excitonic
  absorption in gate-controlled graphene quantum dots},}\ }\href@noop {}
  {\bibfield  {journal} {\bibinfo  {journal} {Phys. Rev. B}\ }\textbf {\bibinfo
  {volume} {82}},\ \bibinfo {pages} {155445} (\bibinfo {year}
  {2010})}\BibitemShut {NoStop}%
\bibitem [{\citenamefont {Jask{\'o}lski}\ \emph {et~al.}(2011)\citenamefont
  {Jask{\'o}lski}, \citenamefont {Ayuela}, \citenamefont {Pelc}, \citenamefont
  {Santos},\ and\ \citenamefont {Chico}}]{Jaskolski2011}%
  \BibitemOpen
  \bibfield  {author} {\bibinfo {author} {\bibfnamefont {W.}~\bibnamefont
  {Jask{\'o}lski}}, \bibinfo {author} {\bibfnamefont {A.}~\bibnamefont
  {Ayuela}}, \bibinfo {author} {\bibfnamefont {M.}~\bibnamefont {Pelc}},
  \bibinfo {author} {\bibfnamefont {H.}~\bibnamefont {Santos}}, \ and\ \bibinfo
  {author} {\bibfnamefont {L.}~\bibnamefont {Chico}},\ }\bibfield  {title}
  {\enquote {\bibinfo {title} {Edge states and flat bands in graphene
  nanoribbons with arbitrary geometries},}\ }\href@noop {} {\bibfield
  {journal} {\bibinfo  {journal} {Phys. Rev. B}\ }\textbf {\bibinfo {volume}
  {83}},\ \bibinfo {pages} {235424} (\bibinfo {year} {2011})}\BibitemShut
  {NoStop}%
\bibitem [{\citenamefont {Cox}\ and\ \citenamefont {Garc{\'\i}a~de
  Abajo}(2014)}]{Cox2014}%
  \BibitemOpen
  \bibfield  {author} {\bibinfo {author} {\bibfnamefont {J.~D.}\ \bibnamefont
  {Cox}}\ and\ \bibinfo {author} {\bibfnamefont {F.~J.}\ \bibnamefont
  {Garc{\'\i}a~de Abajo}},\ }\bibfield  {title} {\enquote {\bibinfo {title}
  {Electrically tunable nonlinear plasmonics in graphene nanoislands},}\
  }\href@noop {} {\bibfield  {journal} {\bibinfo  {journal} {Nat. Comm.}\
  }\textbf {\bibinfo {volume} {5}},\ \bibinfo {pages} {1--8} (\bibinfo {year}
  {2014})}\BibitemShut {NoStop}%
\bibitem [{\citenamefont {Reinhard}, \citenamefont {Brack},\ and\ \citenamefont
  {Genzken}(1990)}]{Reinhard1990}%
  \BibitemOpen
  \bibfield  {author} {\bibinfo {author} {\bibfnamefont {P.-G.}\ \bibnamefont
  {Reinhard}}, \bibinfo {author} {\bibfnamefont {M.}~\bibnamefont {Brack}}, \
  and\ \bibinfo {author} {\bibfnamefont {O.}~\bibnamefont {Genzken}},\
  }\bibfield  {title} {\enquote {\bibinfo {title} {Random-phase approximation
  in a local representation},}\ }\href@noop {} {\bibfield  {journal} {\bibinfo
  {journal} {Phys. Rev. A}\ }\textbf {\bibinfo {volume} {41}},\ \bibinfo
  {pages} {5568} (\bibinfo {year} {1990})}\BibitemShut {NoStop}%
\bibitem [{\citenamefont {Raitza}\ \emph {et~al.}(2012)\citenamefont {Raitza},
  \citenamefont {Reinholz}, \citenamefont {Reinhard}, \citenamefont
  {R{\"o}pke},\ and\ \citenamefont {Broda}}]{Raitza2012}%
  \BibitemOpen
  \bibfield  {author} {\bibinfo {author} {\bibfnamefont {T.}~\bibnamefont
  {Raitza}}, \bibinfo {author} {\bibfnamefont {H.}~\bibnamefont {Reinholz}},
  \bibinfo {author} {\bibfnamefont {P.}~\bibnamefont {Reinhard}}, \bibinfo
  {author} {\bibfnamefont {G.}~\bibnamefont {R{\"o}pke}}, \ and\ \bibinfo
  {author} {\bibfnamefont {I.}~\bibnamefont {Broda}},\ }\bibfield  {title}
  {\enquote {\bibinfo {title} {Spatially resolved collective excitations of
  nano-plasmas via molecular dynamics simulations and fluid dynamics},}\
  }\href@noop {} {\bibfield  {journal} {\bibinfo  {journal} {New J. Phys.}\
  }\textbf {\bibinfo {volume} {14}},\ \bibinfo {pages} {115016} (\bibinfo
  {year} {2012})}\BibitemShut {NoStop}%
\bibitem [{\citenamefont {Townsend}\ and\ \citenamefont
  {Bryant}(2014)}]{Townsend2014}%
  \BibitemOpen
  \bibfield  {author} {\bibinfo {author} {\bibfnamefont {E.}~\bibnamefont
  {Townsend}}\ and\ \bibinfo {author} {\bibfnamefont {G.~W.}\ \bibnamefont
  {Bryant}},\ }\bibfield  {title} {\enquote {\bibinfo {title} {Which resonances
  in small metallic nanoparticles are plasmonic?}}\ }\href@noop {} {\bibfield
  {journal} {\bibinfo  {journal} {J. Opt.}\ }\textbf {\bibinfo {volume} {16}},\
  \bibinfo {pages} {114022} (\bibinfo {year} {2014})}\BibitemShut {NoStop}%
\bibitem [{\citenamefont {Zhang}\ \emph {et~al.}(2017)\citenamefont {Zhang},
  \citenamefont {Bursi}, \citenamefont {Cox}, \citenamefont {Cui},
  \citenamefont {Krauter}, \citenamefont {Alabastri}, \citenamefont
  {Manjavacas}, \citenamefont {Calzolari}, \citenamefont {Corni}, \citenamefont
  {Molinari} \emph {et~al.}}]{Zhang2017}%
  \BibitemOpen
  \bibfield  {author} {\bibinfo {author} {\bibfnamefont {R.}~\bibnamefont
  {Zhang}}, \bibinfo {author} {\bibfnamefont {L.}~\bibnamefont {Bursi}},
  \bibinfo {author} {\bibfnamefont {J.~D.}\ \bibnamefont {Cox}}, \bibinfo
  {author} {\bibfnamefont {Y.}~\bibnamefont {Cui}}, \bibinfo {author}
  {\bibfnamefont {C.~M.}\ \bibnamefont {Krauter}}, \bibinfo {author}
  {\bibfnamefont {A.}~\bibnamefont {Alabastri}}, \bibinfo {author}
  {\bibfnamefont {A.}~\bibnamefont {Manjavacas}}, \bibinfo {author}
  {\bibfnamefont {A.}~\bibnamefont {Calzolari}}, \bibinfo {author}
  {\bibfnamefont {S.}~\bibnamefont {Corni}}, \bibinfo {author} {\bibfnamefont
  {E.}~\bibnamefont {Molinari}},  \emph {et~al.},\ }\bibfield  {title}
  {\enquote {\bibinfo {title} {How to identify plasmons from the optical
  response of nanostructures},}\ }\href@noop {} {\bibfield  {journal} {\bibinfo
   {journal} {ACS Nano}\ }\textbf {\bibinfo {volume} {11}},\ \bibinfo {pages}
  {7321--7335} (\bibinfo {year} {2017})}\BibitemShut {NoStop}%
\bibitem [{\citenamefont {Bernadotte}, \citenamefont {Evers},\ and\
  \citenamefont {Jacob}(2013)}]{Bernadotte2013}%
  \BibitemOpen
  \bibfield  {author} {\bibinfo {author} {\bibfnamefont {S.}~\bibnamefont
  {Bernadotte}}, \bibinfo {author} {\bibfnamefont {F.}~\bibnamefont {Evers}}, \
  and\ \bibinfo {author} {\bibfnamefont {C.~R.}\ \bibnamefont {Jacob}},\
  }\bibfield  {title} {\enquote {\bibinfo {title} {Plasmons in molecules},}\
  }\href@noop {} {\bibfield  {journal} {\bibinfo  {journal} {J. Phys. Chem. C}\
  }\textbf {\bibinfo {volume} {117}},\ \bibinfo {pages} {1863--1878} (\bibinfo
  {year} {2013})}\BibitemShut {NoStop}%
\bibitem [{\citenamefont {Krauter}\ \emph {et~al.}(2015)\citenamefont
  {Krauter}, \citenamefont {Bernadotte}, \citenamefont {Jacob}, \citenamefont
  {Pernpointner},\ and\ \citenamefont {Dreuw}}]{Krauter2015}%
  \BibitemOpen
  \bibfield  {author} {\bibinfo {author} {\bibfnamefont {C.~M.}\ \bibnamefont
  {Krauter}}, \bibinfo {author} {\bibfnamefont {S.}~\bibnamefont {Bernadotte}},
  \bibinfo {author} {\bibfnamefont {C.~R.}\ \bibnamefont {Jacob}}, \bibinfo
  {author} {\bibfnamefont {M.}~\bibnamefont {Pernpointner}}, \ and\ \bibinfo
  {author} {\bibfnamefont {A.}~\bibnamefont {Dreuw}},\ }\bibfield  {title}
  {\enquote {\bibinfo {title} {Identification of plasmons in molecules with
  scaled ab initio approaches},}\ }\href@noop {} {\bibfield  {journal}
  {\bibinfo  {journal} {J. Phys. Chem. C}\ }\textbf {\bibinfo {volume} {119}},\
  \bibinfo {pages} {24564--24573} (\bibinfo {year} {2015})}\BibitemShut
  {NoStop}%
\bibitem [{\citenamefont {Townsend}, \citenamefont {Debrecht},\ and\
  \citenamefont {Bryant}(2015)}]{Townsend2015}%
  \BibitemOpen
  \bibfield  {author} {\bibinfo {author} {\bibfnamefont {E.}~\bibnamefont
  {Townsend}}, \bibinfo {author} {\bibfnamefont {A.}~\bibnamefont {Debrecht}},
  \ and\ \bibinfo {author} {\bibfnamefont {G.~W.}\ \bibnamefont {Bryant}},\
  }\bibfield  {title} {\enquote {\bibinfo {title} {Approaching the quantum
  limit for nanoplasmonics},}\ }\href@noop {} {\bibfield  {journal} {\bibinfo
  {journal} {J. Mat. Res.}\ }\textbf {\bibinfo {volume} {30}},\ \bibinfo
  {pages} {2389--2399} (\bibinfo {year} {2015})}\BibitemShut {NoStop}%
\bibitem [{\citenamefont {Jain}(2014)}]{Jain2014}%
  \BibitemOpen
  \bibfield  {author} {\bibinfo {author} {\bibfnamefont {P.~K.}\ \bibnamefont
  {Jain}},\ }\bibfield  {title} {\enquote {\bibinfo {title} {Plasmon-in-a-box:
  on the physical nature of few-carrier plasmon resonances},}\ }\href@noop {}
  {\bibfield  {journal} {\bibinfo  {journal} {J. Phys. Chem. Lett.}\ }\textbf
  {\bibinfo {volume} {5}},\ \bibinfo {pages} {3112--3119} (\bibinfo {year}
  {2014})}\BibitemShut {NoStop}%
\bibitem [{\citenamefont {Pines}\ and\ \citenamefont {Bohm}(1952)}]{Pines1952}%
  \BibitemOpen
  \bibfield  {author} {\bibinfo {author} {\bibfnamefont {D.}~\bibnamefont
  {Pines}}\ and\ \bibinfo {author} {\bibfnamefont {D.}~\bibnamefont {Bohm}},\
  }\bibfield  {title} {\enquote {\bibinfo {title} {A collective description of
  electron interactions: Ii. collective vs individual particle aspects of the
  interactions},}\ }\href@noop {} {\bibfield  {journal} {\bibinfo  {journal}
  {Phys. Rev.}\ }\textbf {\bibinfo {volume} {85}},\ \bibinfo {pages} {338}
  (\bibinfo {year} {1952})}\BibitemShut {NoStop}%
\bibitem [{\citenamefont {Wallace}(1947)}]{Wallace1947}%
  \BibitemOpen
  \bibfield  {author} {\bibinfo {author} {\bibfnamefont {P.}~\bibnamefont
  {Wallace}},\ }\bibfield  {title} {\enquote {\bibinfo {title} {The band theory
  of graphite},}\ }\href@noop {} {\bibfield  {journal} {\bibinfo  {journal}
  {Phys. Rev.}\ }\textbf {\bibinfo {volume} {71}},\ \bibinfo {pages} {622--634}
  (\bibinfo {year} {1947})}\BibitemShut {NoStop}%
\bibitem [{\citenamefont {Potasz}, \citenamefont
  {G\"u\ifmmode~\mbox{\c{c}}\else \c{c}\fi{}l\"u},\ and\ \citenamefont
  {Hawrylak}(2010)}]{Potasz2010}%
  \BibitemOpen
  \bibfield  {author} {\bibinfo {author} {\bibfnamefont {P.}~\bibnamefont
  {Potasz}}, \bibinfo {author} {\bibfnamefont {A.~D.}\ \bibnamefont
  {G\"u\ifmmode~\mbox{\c{c}}\else \c{c}\fi{}l\"u}}, \ and\ \bibinfo {author}
  {\bibfnamefont {P.}~\bibnamefont {Hawrylak}},\ }\bibfield  {title} {\enquote
  {\bibinfo {title} {Spin and electronic correlations in gated graphene quantum
  rings},}\ }\href {\doibase 10.1103/PhysRevB.82.075425} {\bibfield  {journal}
  {\bibinfo  {journal} {Phys. Rev. B}\ }\textbf {\bibinfo {volume} {82}},\
  \bibinfo {pages} {075425} (\bibinfo {year} {2010})}\BibitemShut {NoStop}%
\end{thebibliography}%

\end{document}